\documentclass{aa}  

\usepackage{natbib}
\usepackage{graphicx}
\usepackage{txfonts}
\usepackage{color}
\definecolor{red}{rgb}{1,0,0}
\definecolor{blue}{rgb}{0,0,1}

\begin{document}
   \title{Imaging the spotty surface of \object{Betelgeuse} in the H band
   }


   \author{X. Haubois\inst{1}
          \and
         G. Perrin\inst{1}
         \and
          S. Lacour\inst{1}
   \and
          T. Verhoelst\inst{2}
   \and
           S. Meimon\inst{3}
   \and
        L. Mugnier\inst{3}
   \and
         E. Thi\'{e}baut\inst{4}
   \and
          J.P. Berger\inst{5}
   \and
          S.T. Ridgway\inst{6}
   \and
          J.D. Monnier\inst{7}
          \and
          R. Millan-Gabet\inst{8}
   \and 
   	W. Traub\inst{9}
          }

   \institute{LESIA, Observatoire de Paris, 5 place Jules Janssen, 92190 Meudon, France
   \\    \email{xavier.haubois@obspm.fr}
  \and   Instituut voor Sterrenkunde, K.U.Leuven, 3001 Leuven, Belgium      
  \and Office National d\'\,Etudes et de Recherches A\'{e}ronautiques, DOTA, 92322 Chatillon, France 
   \and  Centre de Recherche Astrophysique de Lyon, CNRS/UMR 5574, 69561 Saint Genis Laval, France 
 \and Laboratoire d\'\,Astrophysique de Grenoble, CNRS/UMR 5571, 38041 Grenoble, France 
 \and Kitt Peak National Observatory, National Optical Astronomy Observatories,  P.O. Box 26732,  Tucson, AZ 85726-6732 U.S.A. 
 \and University of Michigan, 941 Dennison Building, 500 Church Street, Ann Arbor, MI 48109-1090 USA
  \and Caltech/Michelson Science Center, Pasadena, CA, USA 
 \and Jet Propulsion Laboratory, California Institute of Technology, M/S 301-451, 4800 Oak Grove Dr., Pasadena CA, 91109, USA  }

   \date{Received 21 July 2009 / Accepted 7 October 2009}

 
  \abstract
  {}
   {This paper reports on H-band interferometric observations of \object{Betelgeuse} made at the three-telescope interferometer IOTA. We image \object{Betelgeuse} and its asymmetries to understand the spatial variation of the photosphere, including its diameter, limb darkening, effective temperature, surrounding brightness, and bright (or dark) star spots.}
   {We used different theoretical simulations of the photosphere and dusty environment to model the visibility data.  We made images with parametric modeling and two image reconstruction algorithms: MIRA and WISARD.} 
   {
   We measure an average limb-darkened diameter of 44.28 $\pm$ 0.15\,mas with linear and quadratic models and a Rosseland diameter of 45.03 $\pm$ 0.12\,mas with a MARCS model. These measurements lead us to derive an updated effective temperature of 3600 $\pm$ 66\,K. We detect a fully-resolved environment to which the silicate dust shell is likely to contribute. By using two imaging reconstruction algorithms, we unveiled two bright spots on the surface of \object{Betelgeuse}. One spot has a diameter of about 11 mas and accounts for about 8.5\% of the total flux. The second one is unresolved (diameter $<$ 9 mas) with 4.5\% of the total flux.}
   {Resolved images of \object{Betelgeuse} in the H band are asymmetric at the level of a few percent. The MOLsphere is not detected in this wavelength range. The amount of measured limb-darkening is in good agreement with model predictions. The two spots imaged at the surface of the star are potential signatures of convective cells.}

   \keywords{Convection-techniques: interferometric- stars: fundamental parameters- infrared: stars- stars: individual: \object{Betelgeuse}}

   \maketitle
%

\section{Introduction}
Located in the Orion constellation, \object{Betelgeuse} ($\alpha$ Orionis) is a red supergiant (hereafter RSG) of spectral type M2Iab. It is one of the brightest stars at optical wavelengths and has the second biggest angular diameter \citep[$\sim$43 mas,][]{2004A&A...418..675P} after R Doradus \citep{1997MNRAS.286..957B}. Classified as semi-regular, it shows periodicity in its brightness changes, accompanied or sometimes interrupted by various irregularities \citep{1984LNP...193..336G}. According to a recent reanalysis of a Hipparcos satellite dataset, its distance is now estimated at 197 $\pm$ 45 pc \citep{2008AJ....135.1430H}. The observations up to now have identified at least 7 components of the complex \object{Betelgeuse} atmosphere: two outer shells, a dust environment, a chromosphere, a gaseous envelope, a molecular shell also known as MOLsphere \citep{2000ApJ...538..801T}, and finally the photosphere. Some of these components are not symmetric, and some are overlapping.

\subsection{Two outer shells}
Two shells (S1 and S2) have been observed in absorption in $^{12}$C$^{16}$O and $^{13}$C$^{16}$O at 4.6 $\mu$m by \cite{1979ApJ...233L.135B} but their spatial extent was not directly determined.
Phoenix 4.6 $\mu$m spectra obtained by \cite{2009AIPC.1094..868H} led to an estimation of the size for S1 and S2. An outer radius of $\sim$4.5 and $\sim$7 arcsec were derived, respectively,  although the latter is inconsistent with other measurements made by the CARMA interferometer.

\subsection{A dusty environment}
\label{dustshell}
A shell of dust was first detected with heterodyne interferometry at 11 $\mu$m by \cite{Sutton}. Quite a few estimates of the inner radius of \object{Betelgeuse}'s dust shell have since been made. \cite{Bester1991} used a model with an inner radius of 0.9\,arcsec ($\sim$45\,R$_{\star}$) to explain both their 11\,$\mu$m heterodyne interferometry (ISI) and older speckle observations by \cite{Sutton} and \cite{Howell1981}. In their spatially-resolved mid-infrared (mid-IR) slit spectroscopy, \cite{Sloan1993} find no silicate emission within the central arcsecond around \object{Betelgeuse}\footnote{They also remark that this actually argues against a spherical distribution of the dust.}. \cite{Danchi1994} find from 11.15\,$\mu$m ISI data that the inner radius must be $1.00\pm0.05$\,arcsec, i.e. roughly 50\,R$_{\star}$. But this result disagrees with later findings by \cite{Skinner1997}, who claim an inner radius of not more than 0.5\,arcsec. This environment of dust is also reported by \cite{2007ApJ...670L..21T} as the origin of a visibility drop at short baselines of more than 40\% in new ISI measurements.

\subsection{Chromosphere}
\cite{1975MNRAS.172..277L} interpreted \object{Betelgeuse} spectra and found that  the infrared excess could be explained by a chromospheric emission. More recently, the ultraviolet (UV) observations of the Hubble Space Telescope revealed a spatially resolved chromosphere measuring twice the size of the optical photosphere in the continuum \citep{1996ApJ...463L..29G,1998AJ....116.2501U}.

\subsection{The asymmetric gaseous envelope}
The outer envelope of the O-rich \object{Betelgeuse} \citep[O/C=2.5,][]{1984ApJ...284..223L} has been the subject of extensive studies. VLA observations revealed an extended, cool ($\sim$1000-3000 K) and irregularly shaped gaseous atmosphere of several stellar radii which coexists with the chromospheric gas \citep{1998Natur.392..575L}. More recently \cite{2009arXiv0907.1843K} found evidence for the presence of a bright plume, possibly containing CN, extending in the southwest direction up to a minimum distance of six photospheric radii. Such a plume could be formed either by an enhanced mass loss above a large upwards moving convective cell or by rotation, through the presence of a hot spot at the location of the polar cap of the star.

\subsection{The MOLsphere}
With the Stratoscope II experiment, strong water-vapor bands have been detected in the spectrum of \object{Betelgeuse} \citep{1965ApJ...141..116D} but at that time no explanation on their origin was given. More recently, Tsuji \citep{2000ApJ...538..801T,2000ApJ...540L..99T} proposed the existence of a molecular layer at 0.45 $R_{\star}$ above the photosphere, a MOLsphere, of effective temperature 1500 $\pm$ 500 K  and composed of water vapor and CO. Using infrared interferometry, \cite{2004A&A...418..675P} showed that K, L and 11.15 $\mu$m data could be simultaneously modeled with a gaseous shell of temperature 2055 $\pm$ 25 K located 0.33 $R_{\star}$ above the photosphere while \cite{2006ApJ...637.1040R} locate the water vapor inside the photosphere by analyzing high-resolution spectra. A model based on mid-infrared measurements showed that the MOLsphere could contain not only water vapor but also SiO and Al$_{2}$O$_{3}$ \citep{Verhoelst2006,2007A&A...474..599P}. This latter component is a type of dust condensing at high temperature and on which SiO can be adsorbed. It is a nucleation site at high temperature for silicate dust.  This is a possible scenario to explain dust formation and to connect molecular layers and dusty environments.

\subsection{The convective photosphere}
 \cite{2004A&A...418..675P} modeled the photosphere of \object{Betelgeuse} as a uniform disk of  3641 $\pm$ 53 K temperature with a 42 mas diameter. For some wavelengths, the diameter can vary with time. At 11.15$\mu$m, \cite{2009ApJ...697L.127T} measure a systematic decrease of the  diameter of \object{Betelgeuse} during the period 1993-2009 that might point toward long-term dynamics at its surface. Moreover, the more recent hydro-radiative simulations of RSGs \citep{2002AN....323..213F} foresee the presence of convection cells whose size is comparable to the stellar radius, as suggested by \cite{1975ApJ...195..137S}.
  
\subsection{The spots observed on the surface of \object{Betelgeuse}}
Thanks to the pupil masking technique, \cite{1992MNRAS.257..369W, 1997MNRAS.291..819W}  and \cite{1990MNRAS.245P...7B} have detected the presence of spots on the surface of \object{Betelgeuse} that represent up to 15-20\% of the observed flux in the visible. The most likely origin of these asymmetries is a signature of the convection phenomenon. However, the hypotheses of inhomogeneities in the molecular layers or the presence of a transiting companion are not ruled out \citep{1990MNRAS.245P...7B}. These observed spots are few, 2 or 3, and their characteristrics change with time \citep{1992MNRAS.257..369W}. Confirming previous results, \cite{2000MNRAS.315..635Y} observed spots in the visible but show a fully centro-symmetric picture of \object{Betelgeuse} in the near infrared (at 1.29 $\mu$m). Lately \cite{2007ApJ...670L..21T} also report the observation of an asymmetry at 11.15 $\mu$m located on the southern edge of the disk of \object{Betelgeuse}. In order to deepen our knowledge on RSGs, it is critical to observe spots at high angular resolution. Many unknowns remain such as their size, location, chemical composition, dynamical properties, lifetime and origin.
\vspace{3mm}

To investigate these questions, we report in this paper on interferometric observations obtained with the IONIC 3-telescope beam combiner at the IOTA interferometer in October 2005. Sect.~\ref{obssect} describes the observations and the data reduction. In Sect.~\ref{envsect}, the fully resolved environment is discussed. The physical diameter and limb darkening of \object{Betelgeuse} are investigated in Sect.~\ref{LDsect} . In Sect.~\ref{imasect}, images of \object{Betelgeuse} reconstructed with the MIRA and WISARD algorithms are compared to a parametric fit. Conclusions are gathered in Sect.~\ref{conclusect}.


\section{Observation and Data reduction}
\label{obssect}

\begin{figure}[h!]
\begin{center}
\includegraphics[scale=.5]{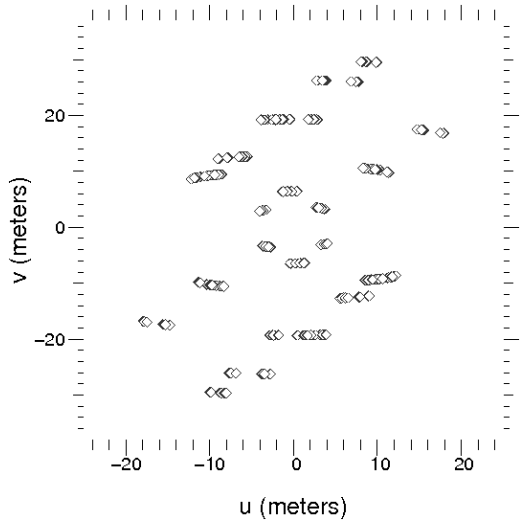}
\end{center}
\caption{uv coverage of the \object{Betelgeuse} observations.}
\label{fig:Bet_uv}
\end{figure}

The data were acquired with the 3-telescope interferometer IOTA \citep[Infrared Optical Telescope Array,][]{2003SPIE.4838...45T} located on Mount Hopkins in Arizona. This instrument has an L shape whose arms are oriented northeast and southeast with lengths of 35 and 15 meters respectively. In order to sample the uv plane, the three telescopes are movable on these arms and can be locked at 17 different locations. The maximum projected baseline is thus 38 meters corresponding to a resolution\footnote{Here, we define the resolution as $\frac{\lambda}{B}$ which is given by the cutoff spatial frequency of the Optical Transfer Function (OTF), $\lambda$ being the observing wavelength and B the projected baseline. We mention that alternative values exist such as the FWHM of the central fringe.} of 9 mas in H band. Light coming from three apertures (siderostats of 0.45 m diameter) is spatially filtered by single-mode fibers which clean the wavefront and remove atmospheric corrugations that affect the contrast \citep{1998SPIE.3350..856C,1998IAUS..189P..18P}. The fringes are scanned with piezo mirrors. The beams are then combined with IONIC \citep{2003SPIE.4838.1099B}. This integrated optics (IO) component combines 3 input beams in a pairwise manner. The combination is coaxial. Light was split in two orthogonal polarization axes with a Wollaston prism. Finally the fringes were imaged with an acquistion camera using a PICNIC detector \citep{2004PASP..116..377P}. We observed $\alpha$ Ori in the H narrow-band filter ($\lambda_{0}$=1.64 $\mu$m, $\Delta\lambda$ =0.10 $\mu$m) during 6 days using 5 different configurations of the interferometer in order to get a rich uv coverage (see Table~\ref{tab:Logob} and Fig.~\ref{fig:Bet_uv}). To calibrate the instrumental transfer function, we observed alternatively $\alpha$ Orionis and a calibrator (HD 36167).

\begin{table*}
\centering
\begin{tabular}{lcccccl}
\hline
Date & Configuration & Maximum projected baseline &  Resolution associated with the\\
&  &  in meters &  maximum projected baseline in mas\\
\hline
October 7, 2005 &	A5-B5-C0     &	6.5 &  51.7
\\
October 8, 2005 &	A5-B15-C0     & 15.2 &  22.1	
\\
October 10, 2005&	A20-B15-C0  & 19.5    & 17.2
\\
October 11, 2005&	A20-B15-C0   &19.6 &  17.1
\\
October 12, 2005&	A25-B15-C0   & 27.2 & 12.4
\\
October 16, 2005&	A30-B15-C15  & 31.2 & 10.8
\\
\hline
\end{tabular}
\caption{Log of the observations. Configurations represent the station locations of the telescopes A,B and C along the northeastern arm (A and C) and southeastern arm (B) of IOTA.}
\label{tab:Logob}
\end{table*}

Data reduction was carried out using an IDL pipeline \citep{2004ApJ...602L..57M,2007ApJ...659..626Z}. For the visibilities, we measured the power spectrum of each interferogram (proportional to the target squared visibility) after correcting for intensity fluctuations and subtracting out bias terms from read noise, residual intensity fluctuations, and photon noise \citep{2003A&A...398..385P}. Then, the data pipeline applies a correction for the variable flux ratios for each baseline by using a flux transfer matrix \citep{1997A&AS..121..379C,2001PASP..113..639M}.  Finally, raw squared visibilities are calibrated using the raw visibilities obtained by the same means on the calibrator stars. 



Closure phases are computed as the phase of the complex triple amplitude, also called bispectrum \citep{1996A&A...306L..13B}. These measurements were recorded when interferences occurred on the three baselines taking advantage of the baseline boot-strapping technique \citep{2005ApOpt..44.5173P}. This method makes it possible to record fringes on a long baseline at low visibility when fringes on the shorter baselines are detected. Pairwise combiners (such as IONIC3) can have a large instrumental offset for the closure phase. This is usually calibrated by using a point-source calibrator. The instrumental closure phase of IONIC3 drifted by less than $1$ degree over many hours, thanks to the compact size of the IO component. For both the squared visibilities and the closure phase, the errors were calculated with the bootstrap resampling method. This technique involves repeated re-estimation of a statistical parameter using random samples with replacement from the original data.

IO combiners  always have some low level of photon crosstalk. Here, the crosstalk phenomenon is defined as the undesired transmission of flux (coherent or not) coming from one telescope to a combiner (therefore baseline) involving two other telescopes. This is a potential source of bias, especially when measuring small visibilities, that needs to be carefully calibrated. We investigated whether the level of visibilities at high frequencies is due to a crosstalk contamination from a shorter baseline. From past crosstalk measurements, we simulated a pessimistic crosstalk where the high contrast obtained on a short baseline contaminates a lower contrast measured on a longer baseline.
The results of these simulations show that the crosstalk cannot account for more than half a percent of the high spatial frequency visibilities we measured, i.e. the maximum potential visibility bias is within error bars. Other observations of Arcturus with IONIC/IOTA \citep{2008A&A...485..561L} revealed a minimum level of 1\% of visibility at high spatial frequency. With a contrast up to $\sim$4.5\%, the visibilities at high spatial frequency observed on \object{Betelgeuse} cannot be explained by a crosstalk effect and we finally conclude they undoubtedly are of astrophysical origin.

\section{Detection of the circumstellar environment of \object{Betelgeuse}}
\label{envsect}
The whole data set comprizes 393 squared visibility and 131 closure phase calibrated measurements. We tried several approaches in order to fit the data. We first compare the data with a simple disk model and then discuss the circumstellar environment and limb darkening.

\subsection{Uniform disk and a fully resolved environment}

\begin{figure}[h!!]
\begin{center}
\includegraphics[scale=.4]{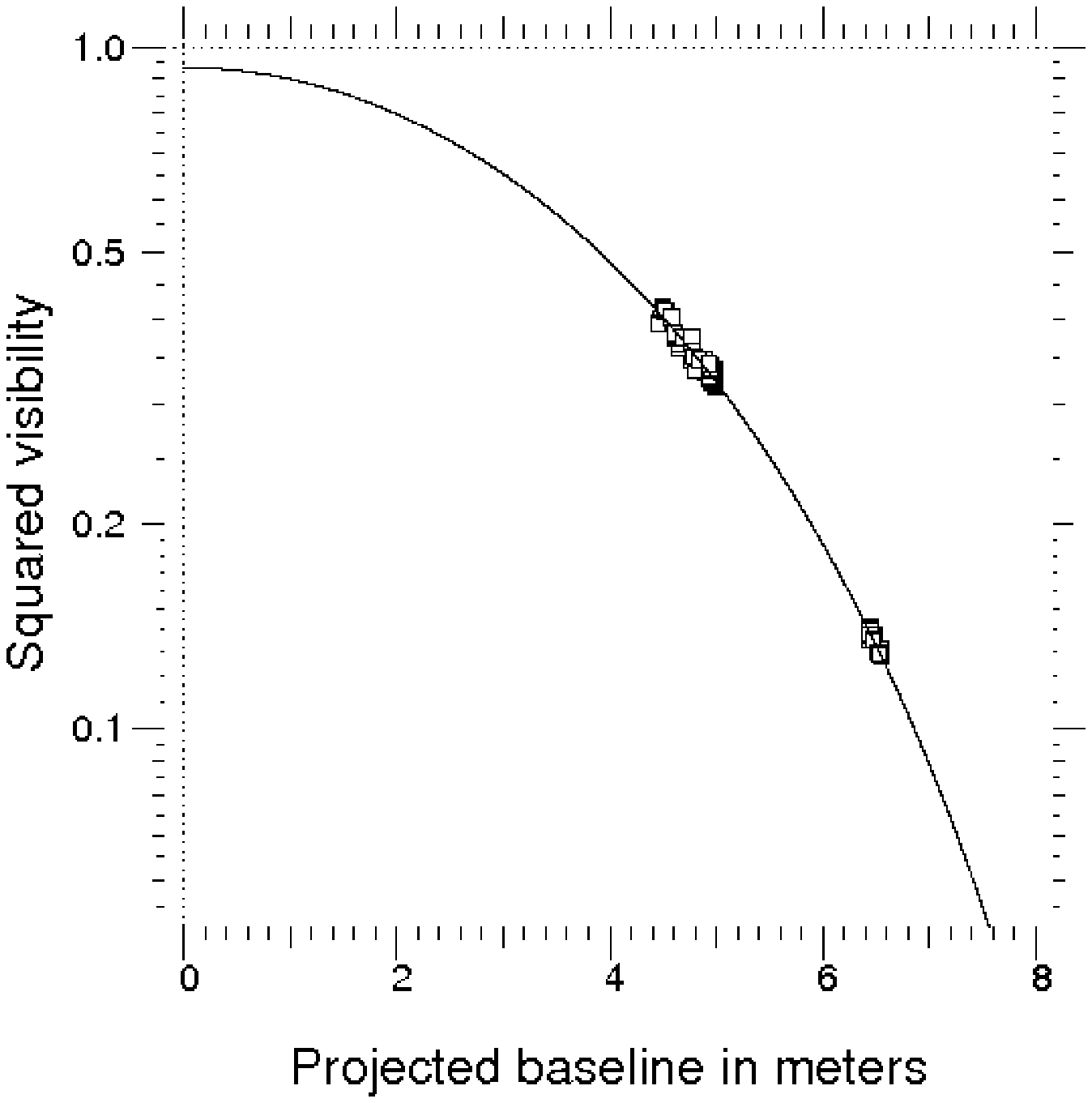} 
\includegraphics[scale=.3]{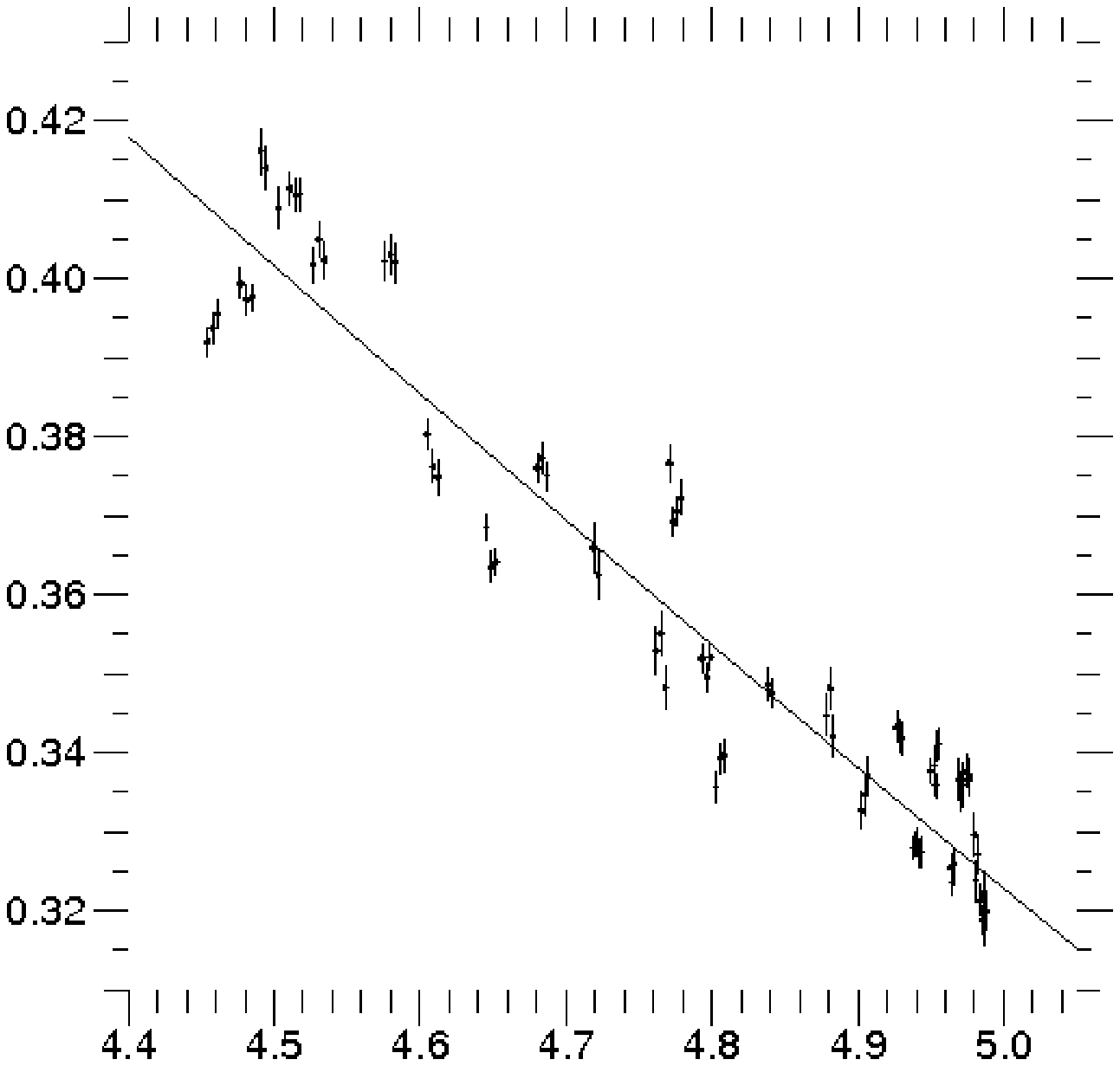} 
\includegraphics[scale=.3]{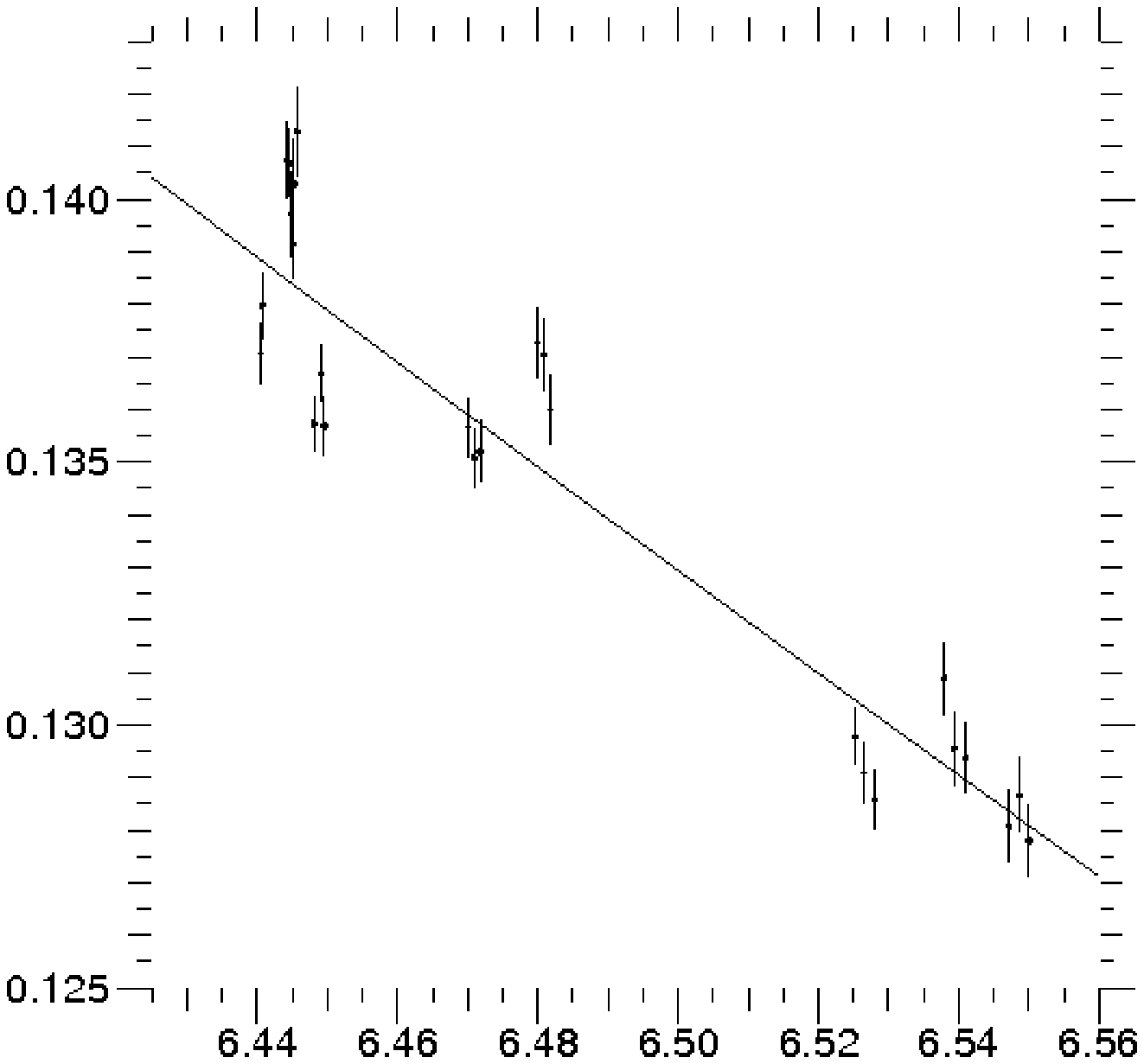} 
\end{center}
\caption{\textbf{Top}: Fitting of a UD plus an environment to the first lobe of \object{Betelgeuse}. \textbf{Bottom}: Enlargements of the two parts of the first lobe. Squared visibility at null baseline is equal to  0.93 (meaning 0.96 in visibility).}
\label{fig:Bet_DU2}
\end{figure}

A first model that can be fit to interferometric data is the uniform disk (UD). It has only one parameter, the UD diameter $\phi_{UD}$. It is used to give a first approximation to the object's angular size. Only first lobe data are used. Higher order lobes contain information on limb darkening and on smaller details.
A  loss of visibiity at low spatial frequencies (i.e. at short baselines) needs to be modeled by scaling the visibility function. Thus the squared visibility is

\begin{equation}
  \begin{array}{l}  
   | V(B,\phi)  |^{2} =\displaystyle  \Big| \omega \times  \int_{\lambda_{0}-\frac{\Delta \lambda}{2}}^{\lambda_{0}+\frac{\Delta \lambda}{2}}  \Big(\frac{2 J_{1} (\pi \phi B_{\perp} / \lambda)} {\pi \phi B_{\perp} / \lambda}\Big) d\lambda \Big|^{2}
   \end{array}
\label{UDe}
\end{equation}
where $B_{\perp} = \sqrt{u^{2}+v^{2}}$ and $J_{1}$ is the second Bessel function of the first kind. The constant factor $\omega$ represents the loss of contrast at null baseline due to the presence of a fully resolved environment, i.e., circumstellar light, from a region outside the stellar disk.  Consequently, the visibility of the central object at the null baseline is not equal to one anymore but to this scaling factor. The fit is shown in Fig.~\ref{fig:Bet_DU2}.

The second part of the first lobe dataset is well reproduced by this model. Consequently, the reduced $\chi^{2}$ changes from 42 (for one UD) to 16 (UD + environment). This model gives a UD diameter of 42.41  $\pm$ 0.07 mas and a coefficient $\omega$ of 0.96 $\pm$ 0.01, meaning that the environment is responsible for about  4 \% of the total flux. That UD diameter is about one mas smaller than the previous measurement of 43.26 mas made by \cite{2004A&A...418..675P} based on data in K band. 

A loss of constrast can also exist when one observes an object so bright that the detector works in a non-linear regime or even saturates. This case was possible given the high flux of \object{Betelgeuse} in H band ($m_{H}$=-3). Although we used diaphragms to decrease the stellar flux when we acquired the data, it is conceivable that we approached the non-linear regime of the detector due to variability in the observing conditions. After a visual analysis of the interferograms and the study of a photometric estimator, we concluded that our data do not suffer from non-linearity or saturation from the detector. Moreover, for all the observed stars during that campaign, no other has shown a similar effect. Thus, the effect does not have an instrumental origin.

The necessity to account for a visibility scaling factor is hence the first evidence of the presence of a fully resolved environment in the near infrared. Even the shortest baseline data do not appear to be contaminated by correlated flux from this extended component at the level of the visibility accuracy. 
A rough estimate of the minimum size of this environment can be established by considering no departure from the uniform disk model is detected at short baseline. Assuming simple models for the environment like a Gaussian distribution or a uniform disk distribution and a maximum contribution of 1\% to the visibility residuals, minimum sizes of around 100 and 400 \,mas are respectively derived. The size of the environment is therefore probably at least a few hundred mas. The visibility drop could stem from an extended environment in the upper atmosphere or to a dusty shell. We study these hypotheses in the following.

\subsection{Extended emission}

Extended broadband emission at a scale of $\sim$100\,mas and more has been detected around \object{Betelgeuse} in mid-IR, radio and at UV wavelengths. The mid-IR  is caused by thermal dust emission \citep[see e.g.][]{Danchi1994}. The radio emission mostly arises from free-free processes in the gaseous envelope \citep{Harper2001}. The UV emission is a result of a combination of scattered and fluorescent line photons, electron collisional excitation, and bound-free recombination in the chromosphere and/or ionized wind \citep{Harper2001,Lobel2001}. Although the H band covers an opacity minimum of the combined free-free and bound-free emission from both chromosphere and (partly) ionized wind in the gaseous envelope, we can not exclude these emission processes as the origin of the observed environment without detailed calculations based on the models which were constructed to explain the UV and radio observations \citep[e.g.]{Harper2001}. However, as is shown in this section, scattered light on the inner edge of the dust shell provides a natural explanation of the observed loss of fringe contrast. The calculation of a possible gaseous emission component at these wavelengths is left to a future paper.

\subsection{Dust shell model and scattered light contribution}

\begin{figure}
\centering
  \resizebox{\hsize}{!}{\includegraphics{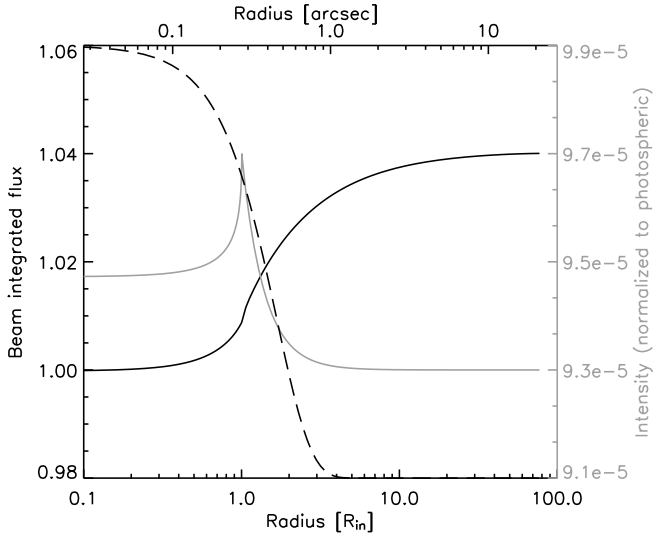}}
  \caption{Amount of beam integrated flux and dust shell intensity compared to the fiber profile. {\sl Black line}: beam integrated flux, normalized to the photospheric flux, as a function of beam radius, either in units of dust shell inner radius (bottom axis) or in arcseconds assuming  a 13\,R$_{\star}$ dust shell inner radius (top axis). {\sl Grey line}: the intensity profile of the dust shell, normalized to the central photospheric intensity. The dashed line represents the Gaussian fiber profile.} 
  \label{fig:scattered}
\end{figure}

The dust shell was modeled using the proprietary spherical radiative
transfer code {\sc modust} \citep{Bouwman2000,Bouwman2001}. Under the
constraint of radiative equilibrium, this code solves the
monochromatic radiative transfer equation from UV/optical to
millimeter wavelengths using a Feautrier type solution method
\citep{Feautrier1964, Mihalas1978}. The code permits several
different dust components of various grain sizes and shapes, each with
their own temperature distribution. In \object{Betelgeuse}, the main spectral
dust feature is that of amorphous Magnesium-Iron silicates. Input parameters were derived by modeling the observed IR spectrum in \cite{Verhoelst2006}. This kind of modeling of optically-thin dust shells, which is based on (a limited part of) the IR spectrum and on the optical obscuration, is slightly degenerate between inner radius and mass-loss rate, especially when the exact dust composition is not known.\\
Regardless of the exact inner radius, we find that any dust shell model explaining the observed silicate feature at mir-IR wavelengths produces roughly 4\% scattered stellar light in the H\,band. The crucial question is: which fraction of this scattered light makes it into the IONIC fibers (with a field of view of 750 mas diameter) ? This depends solely on the assumed inner radius. Fig.~\ref{fig:scattered} shows the amount of excess making it into the fiber for a model with a 13\,R$_{\star}$ inner radius. Squared and Gaussian profiles have been choosen to simulate the fiber field of view. We used diaphragms to avoid saturation of the detector during the observations. The diaphragms were located just before the fiber injection units. We thus take into account that the width of the fiber field of view was changing depending on the size of the diaphragm. For the short baselines where the fully resolved flux has been detected, the same diaphragm was employed. So we simulate our profiles for two realistic widths of the fiber field of view, with and without diaphragm: respectively 2.85 and 0.76 arcseconds. In both cases, we expect a roughly 2\% fully-resolved flux. Our observations may be compatible with a smaller inner radius, resulting in more scattered light making it into the fiber.


%

\section{Limb-darkening and physical diameter}
\label{LDsect}

The limb darkening effect is a gradual decrease of surface brightness from the center to the stellar limb caused by the decrease of temperature with altitude: the optical depth of 1 is reached for lower altitudes at the center of the stellar disk than at the edge. Taking this effect into account is mandatory for an accurate measurement of the diameter of the photosphere. To do so, we used parametric models \citep[see][]{2004A&A...428.1001C,1997A&A...327..199H} and a physical model derived from a MARCS simulation of the \object{Betelgeuse} atmosphere.

\subsection{Parametric models}
In order to compare different observations, one needs standard models of the stellar intensity distribution. The simplest ones are the linear and quadratic limb darkening laws. The visibility of a point-symmetric object is a Hankel transform (whose kernel is a Bessel function) normalized by the total spatial intensity distribution. For the monochromatic case, the visibility is 

\begin{equation}
  \begin{array}{l}
V_{\lambda}(r) = \displaystyle\frac{ \left| \int_{0}^{\infty}{I(r,\lambda) J_{0}(2\pi \displaystyle\frac{B_{\perp}}{\lambda} r) r dr}\right|}{\left|{\int_{0}^{\infty}}{I(r,\lambda)rdr}\right|}
\end{array}
\end{equation}
where $B_{\perp}$ is the projected baseline of the interferometer, r the distance to the star center, $I(r,\lambda)$ the spatial intensity distribution at the observing wavelength $\lambda$ and $J_{0}$ is the first Bessel function of the first kind. We present now the calculation of the visibility for a quadratically limb darkened disk. For an achromatic quadratic limb darkening, the spatial intensity distribution is a function of $\mu$, the cosine of the angle between the radius vector and the line of sight. It can be written as follows:

\begin{equation}
  \begin{array}{l}
I  (\mu) = 1 - A (1-\mu)-B(1-\mu)^{2}
\end{array}
\end{equation}
with $\mu = \sqrt{1 - (\frac{2r}{\phi_{LD}})^{2}}$  where $\phi_{LD}$ is the limb darkened disk diameter. With B=0, one gets the linear limb darkening case. Substituting the intensity equation in the visibility expression, the Mellin transform appears. For the general case of a power law of index $\delta$, $I (\mu)=\mu^{\delta}$, this can be written as follows \citep{1997A&A...327..199H}:
\begin{equation}
  \begin{array}{l}
V_{\nu}(x) =   \displaystyle2 \nu \left| \int_{0}^{1}(1 - \mu^{2} )^{\nu -1} J_{0}(x \mu) \mu d\mu \right| = \Gamma (\nu+1) \times \frac{|J_{\nu}(x)|}{(x/2)^{\nu}}
\end{array}
\end{equation}

where $J_{\nu}$ is the Bessel function of the first kind of index $\nu$, defined as $\nu =\delta/2 + 1$. 

\begin{figure}[h!]
\begin{center}
\includegraphics[scale=.3]{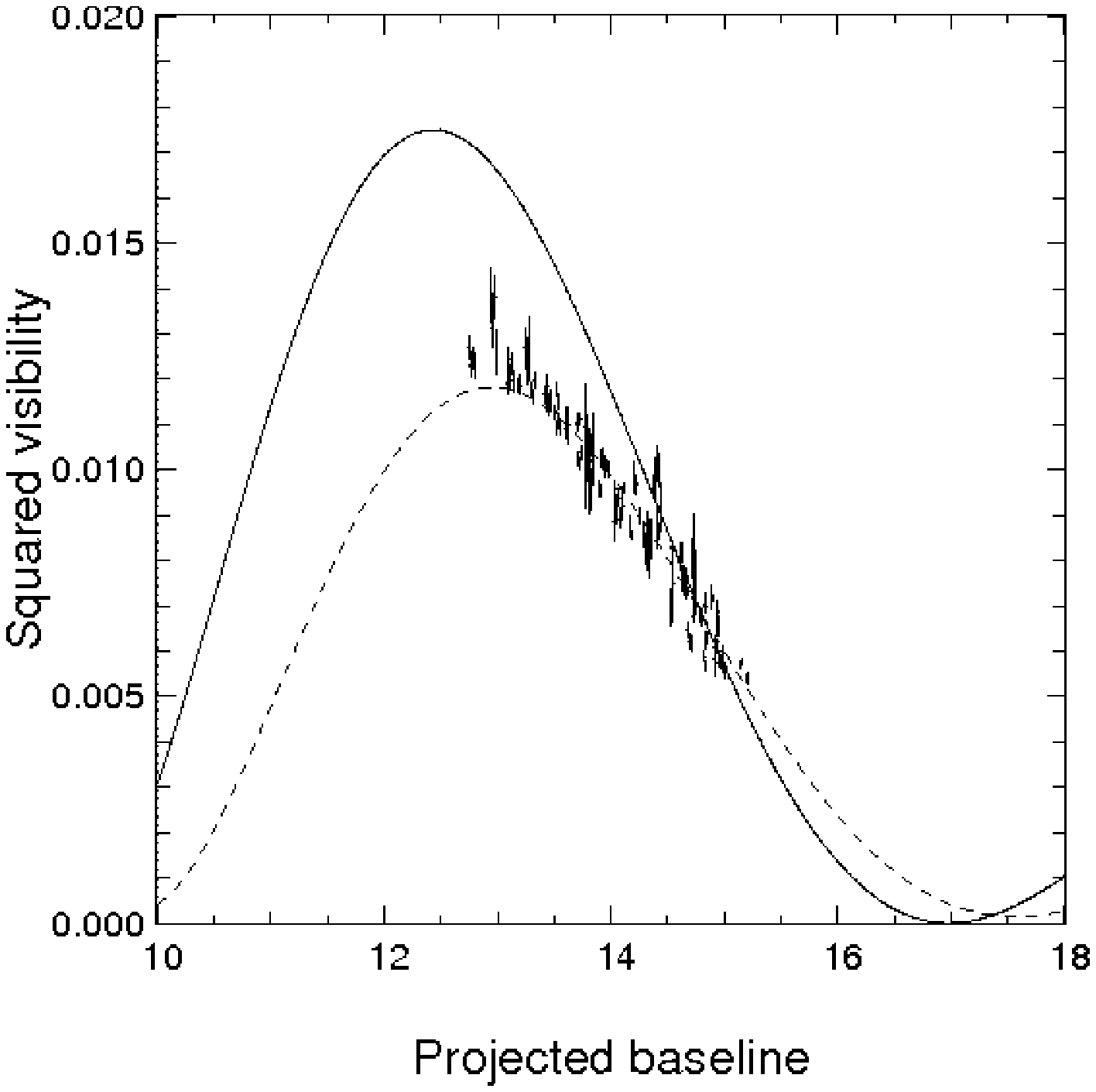}
\includegraphics[scale=.3]{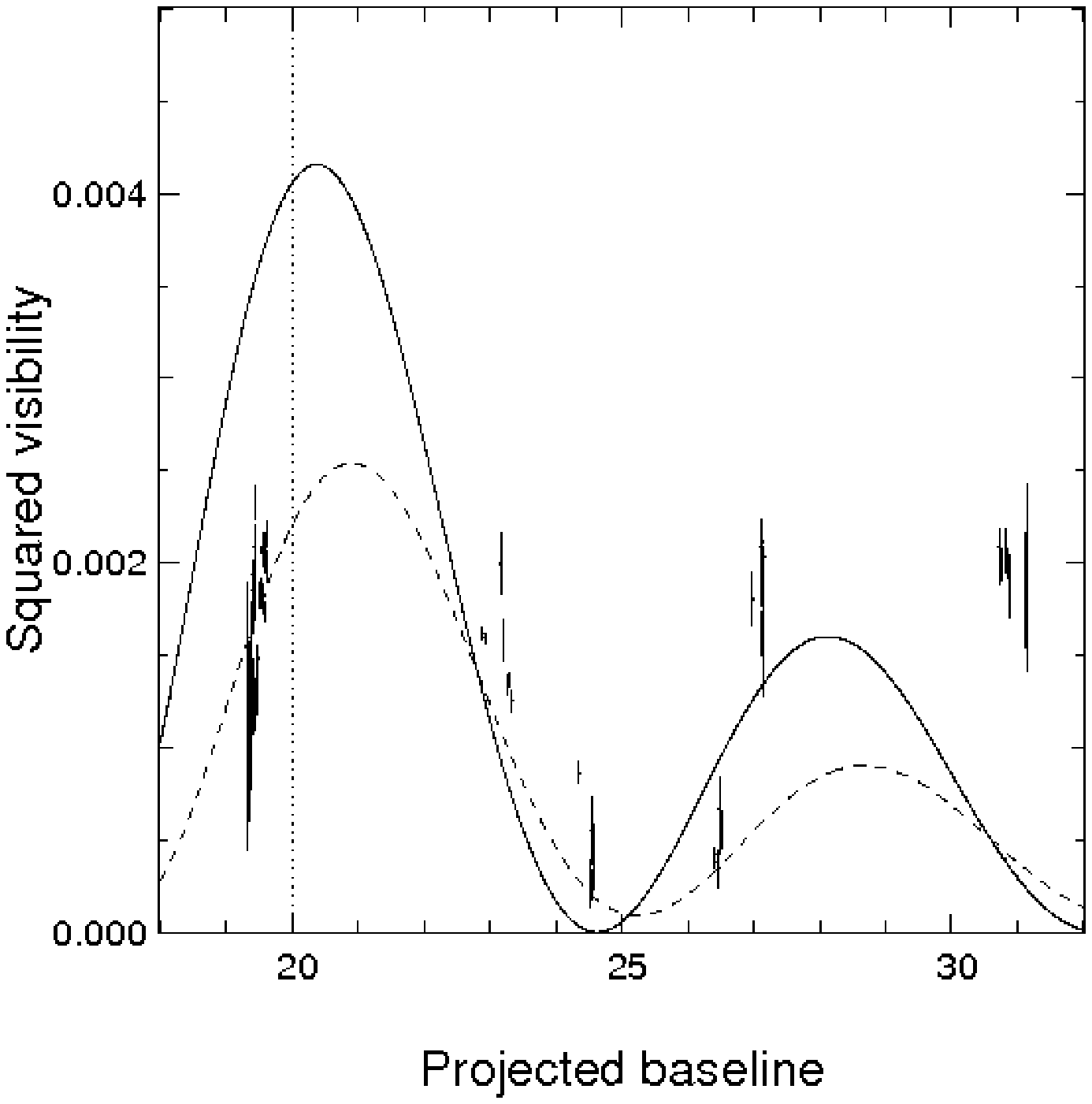}
\end{center}
\caption{Squared visibility data fitted by the best limb darkening model (dotted line). To improve the modeling, we should reproduce the excess of contrast at high frequencies. The squared visibility of a UD is drawn in full line for comparison.}
\label{fig:DAq}
\end{figure}

The expression for the visibility of a quadratic limb-darkening is thus finally:

\begin{equation}
  \begin{array}{l}
V_{\lambda}(z) =\displaystyle\frac{\alpha\displaystyle \frac{|J_{1}(z)|}{z} + \beta \sqrt{\displaystyle\frac{\pi}{2}}\displaystyle \frac{|J_{3/2}(z)|}{z^{3/2}} +2 \gamma \displaystyle\frac{|J_{2}(z)|}{z^{2}}   }{\displaystyle\frac{\alpha}{2} + \displaystyle\frac{\beta}{3} + \frac{\gamma}{4}}
\end{array}
\end{equation}
with $\alpha = 1-A-B$, $\beta = A+2B$, $\gamma=-B$  and $ z = \pi \phi_{LD} \frac{B_{\perp}}{\lambda}$. This model and the linear one have been squared and fit to the data (Fig.~\ref{fig:DAq}). The results of this fitting up to the second lobe are summarized in Table~\ref{tab:DAsum}.

The third and fourth visibility lobes of \object{Betelgeuse} clearly contain high spatial frequency information more complex than just limb darkening and, up to this point, have been excluded from the fits. The linear limb darkening coefficient A=0.43 for a diameter of 44.28 $\pm$ 0.15\,mas can be compared with the estimation by  \cite{2004A&A...418..675P} in K band who found A=0.09 for a smaller diameter of $\sim$43.6 mas. Those measurements obtained in two different bands do not correspond to the same extension of the photosphere. That could explain the difference in diameter. 
On the theoretical side, \cite{1979A&AS...36..411M} measured A=0.52 at 1.2 $\mu$m by fitting a simulated intensity profile of \object{Betelgeuse} with the following parameters: $T_{eff}$ = 3750 K, log g = 1.5 and a solar composition. More recently \cite{1993AJ....106.2096V}  predicted A=0.49 for a limb darkening in J band with $T_{eff}$ = 3500 K, log g = 0.5 and a solar composition. Given that limb darkening is probably smaller in the H band than in the J band, the agreement with our results is quite good.

\subsection{The MARCS modeling of the photosphere}

\begin{figure}[h!]
\begin{center}
\includegraphics[scale=.45]{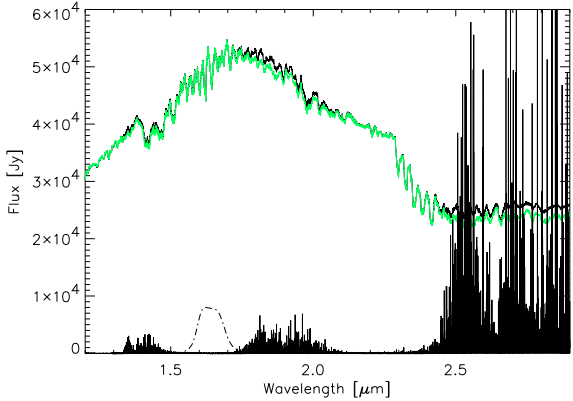}
\end{center}
\caption{The photospheric spectrum (MARCS model) is plotted in black. The photosphere seen through the water layer suggested by \cite{Verhoelst2006} is in green. The optical depth of that layer (times 10$^4$ for visibility) is represented by the vertical lines, and the dashed line
shows that the H-band filter is unaffected by circumstellar water opacity.}
\label{fig:spec}
\end{figure}

\begin{figure}[h!]
\begin{center}
\includegraphics[scale=.45]{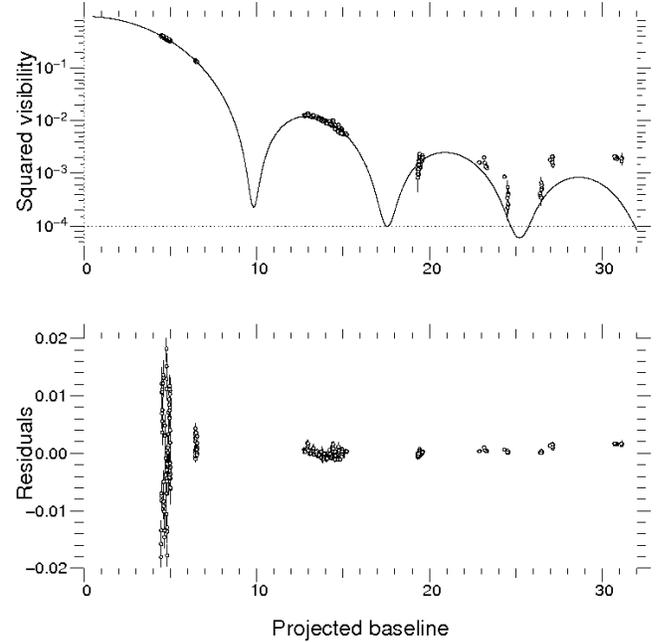}
\end{center}
\caption{\textbf{Upper panel}: MARCS H filter squared visibility (in solid line) fitted to the data (dots) with diameter and resolved flux ratio as free variables. \textbf{Lower panel}: residuals of the fit in the upper panel. We clearly see that this MARCS modeling is convincing up to the third lobe.}
\label{fig:MARCS}
\end{figure}

To model the photosphere of this supergiant, we use the SOSMARCS code, version May 1998, described in \cite{Verhoelst2006} as developed by \cite{1975A&A....42..407G} at Uppsala University, Sweden \citep{1992A&A...256..551P,1993ApJ...418..812P}. This code was specifically developed for the modeling of cool evolved stars, i.e. a lot of effort was put into the molecular opacities, and they allow for the computation of radiative transfer in a spherical geometry. A spectrum derived from this modeling is shown in Fig.~\ref{fig:spec}. In that figure we note that the H narrow-band filter clearly does not see much water. Also CO and SiO molecules have been found in the MolSpheres around evolved stars, but neither molecule shows strong opacity in the H band. More interesting is the suggested presence of CN in a plume originating from the surface of \object{Betelgeuse} \citep{2009arXiv0907.1843K}. CN was not detected in these latter H-band data even if it does have an absorption band with some opacity across these wavelengths (see Fig.\,9 in that paper). As their dynamic range was much higher than the one obtained in the observations analyzed here, CN is therefore not a concern for the results presented in this paper. From the brightness distributions produced by this model, we can compute H-band visibilities, to be compared with the IOTA observations. The result is shown in Fig.~\ref{fig:MARCS}.

The fit of the MARCS model leads to a photospheric limb-darkened diameter of 46.74 $\pm$ 0.12\,mas (see Table~\ref{tab:DAsum}). The  resolved flux ratio is estimated to $\sim$5$\%$. A Rosseland diameter (given by the distance from the star's center at which the Rosseland optical depth equals unity, $\tau_{Ross}$ = 1) of 45.03 $\pm$ 0.12\,mas can be deduced.

\subsection{Effective temperature}
The effective temperature computed in Perrin et al. (2004) can be updated with these new diameter measurements. Following \cite{1992AJ....104.1982D}, a visual extinction of 0.5 was adopted in this calculation. We used the bolometric flux reported in Perrin et al. (2004): 111.67 $\pm$ 6.49 10$^{-13}$Wcm$^{-2}$. Using these values, an average diameter of 44.28 $\pm$ 0.15\,mas for the limb darkening model and a Rosseland diameter of 45.03 $\pm$ 0.12\,mas for the MARCS model lead to effective temperatures of 3612 $\pm$ 53\,K and 3587 $\pm$ 53\,K respectively. We propose the updated slightly lower value of 3600 $\pm$ 66\,K for the effective temperature of \object{Betelgeuse}. The effective temperature has no meaning about the local temperature at the surface of the star. It is a radiometric quantity that characterizes the mean surface brightness of the star. The spots imaged with IONIC on IOTA (Sect.~\ref{anarecons}) induce local variations of the surface brightness. It is interesting to compute how the effective temperature is sensitive to these spots. From our measurements, we deduce that the spots contribute to 2\% of the effective temperature, the same order as the uncertainty on the effective temperature. Current effective temperature measurements are therefore not accurate enough to detect spot fluctuations.


\subsection{Discussion}

\begin{table}[h!]
\centering
\begin{tabular}{lccccccl}
\hline
Model&  $\phi_{LD}$ & Limb darkening & Reduced\\  
  & & parameter(s )& $\chi^{2}$ \\ \hline
Linear  & 44.28 $\pm$ 0.15  & 0.43 $\pm$ 0.03  &  11 \\
\hline
Quadratic  & 44.31 $\pm$ 0.12  &  0.42 $\pm$ 0.03 et 0.02 $\pm$ 0.03  & 11 \\ 
\hline
MARCS  & 46.74 $\pm$ 0.12  &      & 12 \\ 
     \hline
\end{tabular}
\caption{\label{tab:DAsum}Results of fitting of the limb darkening models.}
\end{table}


 The residuals show that there is a discrepancy at small and large baselines. At short baselines, we see that large residuals scatter in a symmetric manner around zero. This variation of visibility at these baselines is likely due to the effect of convection on the diameter measurement at different position angles. The impact of convection on our interferometric measurements is being investigated by some co-authors of the present paper \citep{Chiavassa et al. 2010}. It is remarkable that visibility data can be very well fit by a limb-darkened disk and that no MOLsphere is required, consistent with earlier studies in K \citep{2004A&A...418..675P}. At large baselines, we see once again that the model cannot reproduce the observed high levels of visibility. 

The use of a limb-darkening disk model greatly decreases the reduced $\chi^2$ compared to a uniform disk model. A simple linear law seems to model this effect up to the third lobe. Beyond that limit, the data show a great complexity which is not reproduced by the simpler models. Contrary to AMBER K-band observations made in January 2008 and recently published in \cite{2009arXiv0906.4792O} where the data points are scattered around a limb darkening model, the present observations reveal significant departure from symmetry.
 
  All limb darkening models that we used (linear, quadratic and MARCS model) hint at the presence of additional asymmetric structures suggested by the excess of contrast at high frequency and are beyond the scope of the limb-darkening models. In order to investigate this effect, we analyzed the closure phase measurements to determine the nature of asymmetric structures.

\section{Image reconstruction and asymmetries}
\label{imasect} 
\begin{figure}[h!]
\begin{center}
\includegraphics[scale=.3]{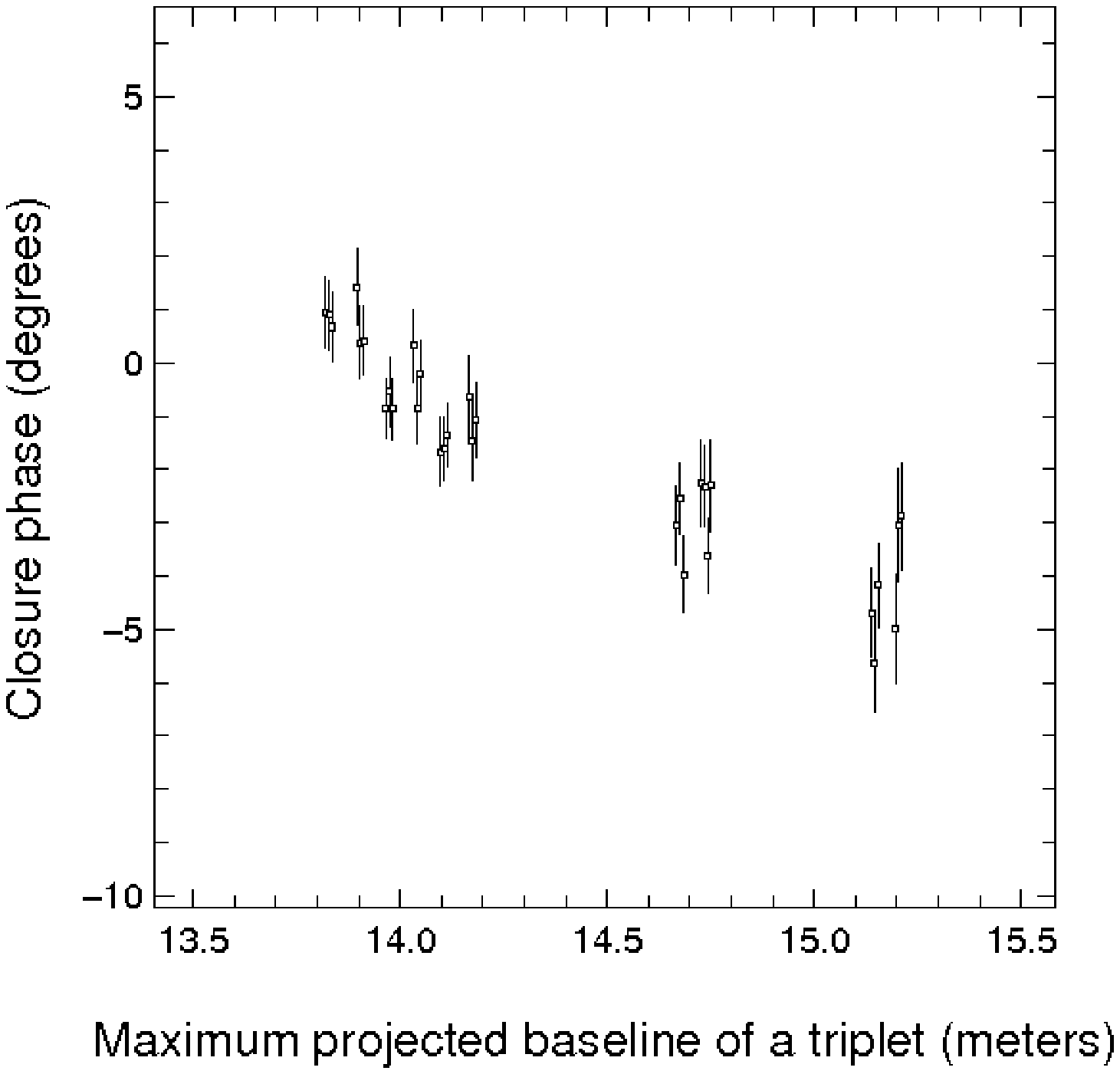}
\includegraphics[scale=.3]{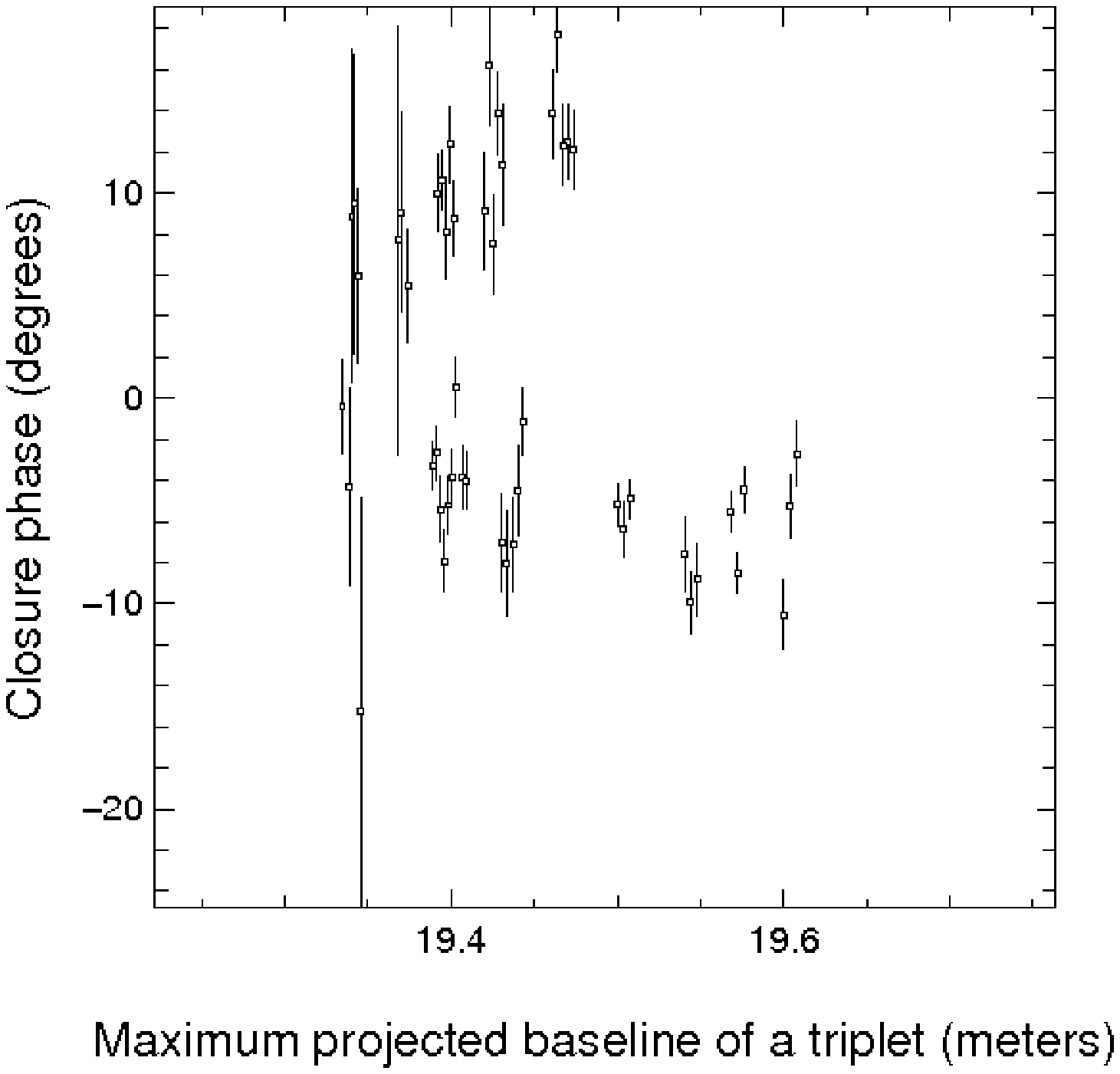}
\end{center}
\caption{Closure phase measurements plotted versus two ranges of the triplet maximum projected baseline.}
\label{fig:cloture}
\end{figure}

The dataset comprises 131 closure phase measurements. The first thing to note is that closure phases are not null over the whole dataset showing departure from point symmetry over \object{Betelgeuse}'s stellar disk. Structures appear in closure phases when plotted versus triplet maximum projected baseline (Fig.~\ref{fig:cloture}).

 With the lack of spatial frequency coverage of near-infrared interferometric observations, parametric modeling with simple geometrical descriptions turns out to be inefficient. Even with a few hundred visibility and closure phase measurements, it is difficult to distinguish basic models because of the presence of a great number of local minima. Thus the parametric approach is not a good way to develop a complex model. In the last decade, various teams have therefore undertaken the creation of image reconstruction algorithms adapted to interferometric data in the near infrared. In this paper we used two of them: MIRA and WISARD. Both are presented and compared in \cite{article_Mugnier} and in \cite{2008A&A...485..561L}.
The major advantage of such methods is that they can reveal structures by reconstructing an image which fits the data and which is also close to an a priori image of the object. The balance between the regularization of the reconstructed image to the a priori image and the data fitting is allowed by setting a hyperparameter $\mu$:

\begin{equation}
\begin{array}{l}
J_{total} = J_{data} + \mu J_{a priori} 
\end{array}
\label{crit}
\end{equation}
where $J_{data}$, $J_{a priori}$ are the criteria that quantify how the reconstructed image fits the data and how it is close to the a priori image.  $J_{total}$ is the final quantity to minimize.

In the present work, we chose as an a priori object the best limb-darkened fit. Our prior is thus a purely symmetric limb darkened disk of 44.28\,mas diameter surrounded by an environment as found in section~\ref{LDsect}. In order to compare the two methods we made sure to use the same field of view, same sampling and same a priori image. Both use a quadratic regularization meaning that strong intensity gradients between the a priori image and the reconstructed image are quadratically discriminated against.

\subsection{MIRA}
	MIRA \citep[Multi-aperture Image Reconstruction Algorithm][]{2008SPIE.7013E..43T} is a method that iteratively minimizes a criterion to reconstruct the image. To minimize the criterion, MIRA uses the optimization method VMLMB \citep{2002SPIE.4847..174T}, a limited variable metric algorithm which accounts for parameter bounds. This last feature is used to enforce positivity of the solution. To avoid falling in local minima, the iteration starts with a strong regularization to the a priori image. Then the weight of the regularization decreases to favor the data fit (i.e. $\mu$ decreases). In terms of reduced $\chi^2$, the best reconstruction achieved gave a $\chi^{2}$ value of 5.7. The reconstructed image is shown in Fig.~\ref{fig:recons}.



\subsection{WISARD}
WISARD \citep[Weak-phase Interferometric Sample Alternating Reconstruction Device, ][]{SergephD,2008JOSAA..26..108M}  has a more explicit approach to the missing phase information issue. Using a self calibration method inspired from radio interferometry \citep{1981MNRAS.196.1067C}, it aims at reconstructing an image and also finding the missing phase data. WISARD uses an appropriate approximation of the specific noise encountered in infrared interferometric data, which favors the convergence toward a global minimum - see \cite{2005JOSAA..22.2348M} for details.

\begin{figure}[h!]
\begin{center}
\includegraphics[scale=.3]{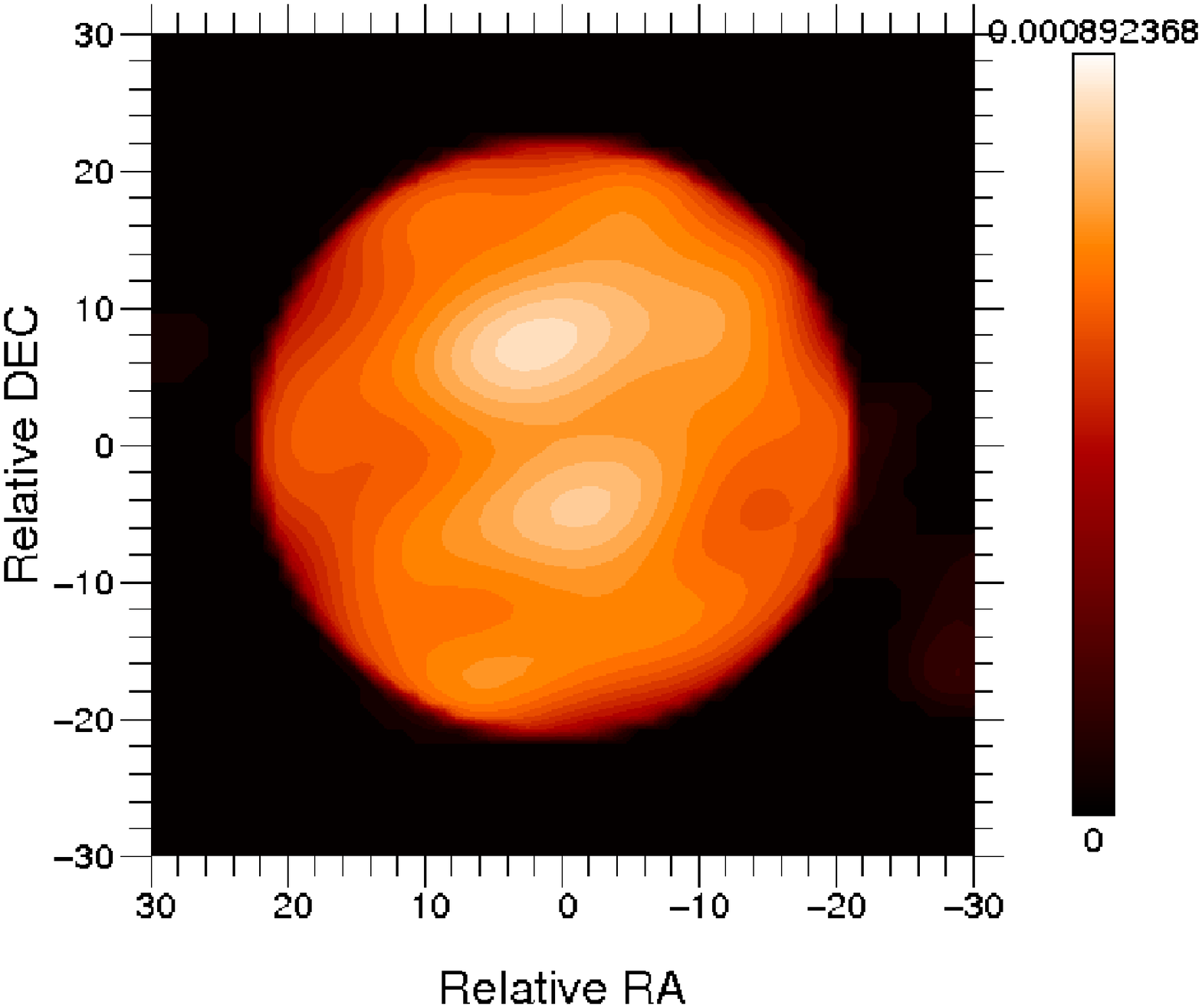}
\includegraphics[scale=.3]{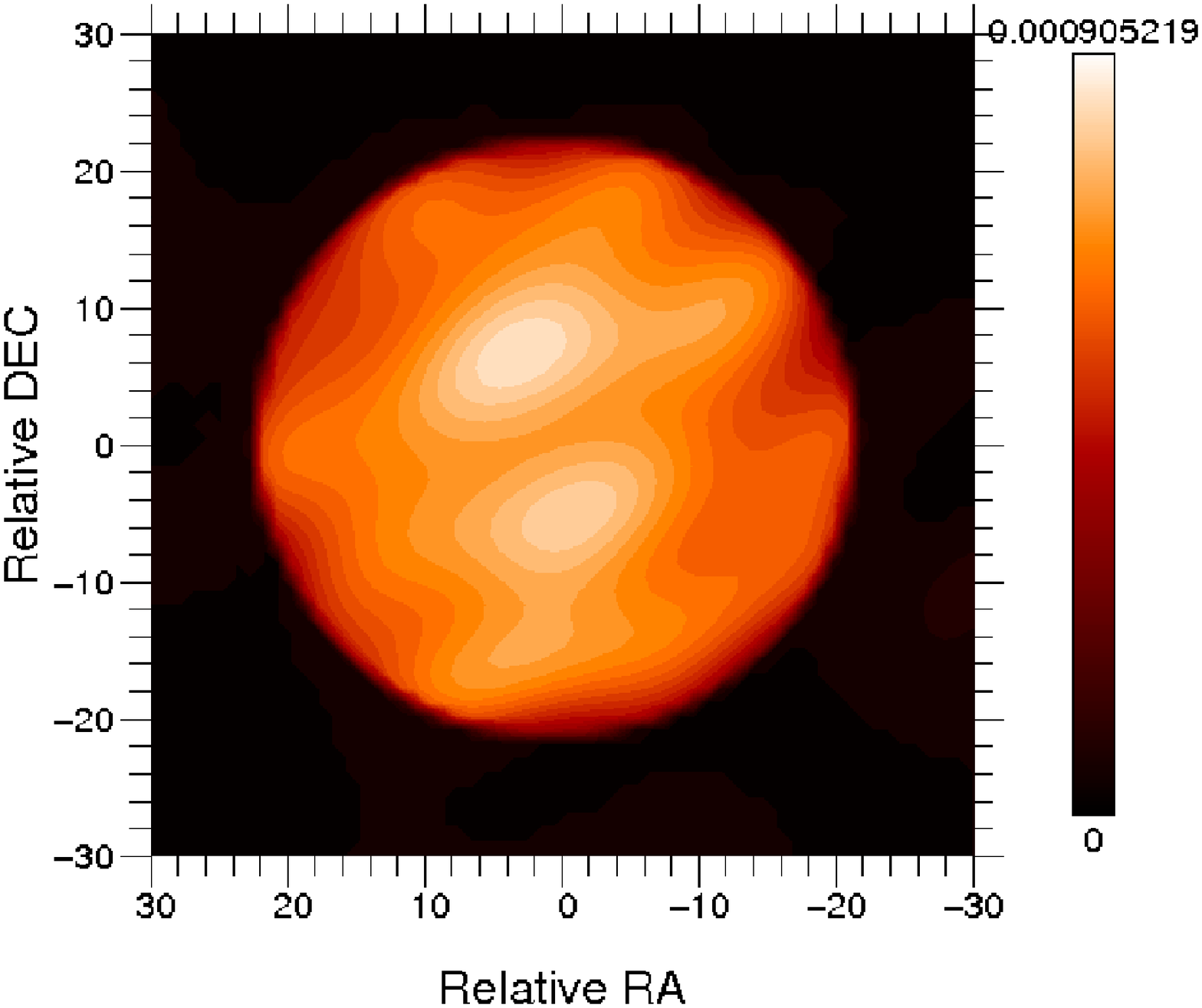}
\includegraphics[scale=.45]{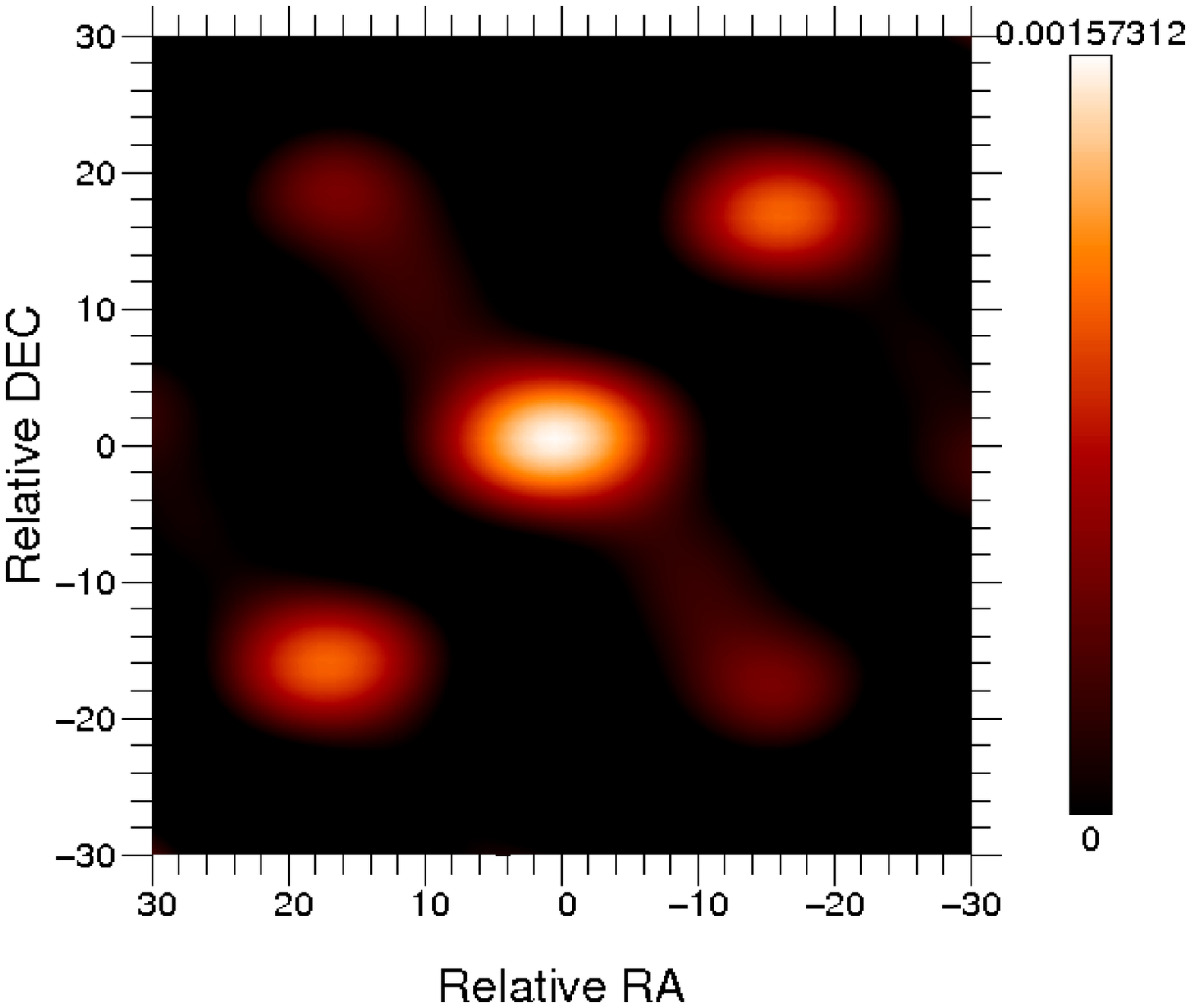}
\end{center}
\caption{\textbf{Upper window}: Contour image reconstruction from MIRA. \textbf{Middle window}: Contour image reconstruction from WISARD of \object{Betelgeuse} in a 60 mas field. Both images were reconstructed with the same a priori object and the same type of regularization. \textbf{Lower window}: Image of the dirty beam, i.e., the interferometer's response to a point source, given our uv plane coverage.}
\label{fig:recons}
\end{figure}

 For the best reconstruction obtained with WISARD, the minimum value of this criterion corresponded to a reduced $\chi^{2}$ of 4.7. The reconstructed image is presented in Fig.~\ref{fig:recons}.


\subsection{Analysis of the reconstructed images}
\label{anarecons}
The first thing to note is the impressive similarity between the two images given the complexity of the data. The two algorithms both reveal two spots (named T1 and T2) of unequal brightness and which are roughly located at the same positions (Table~\ref{tab:2_spots}).  Apart from T1 and T2, the reconstructed images look very similar. The fact that two independent methods give similar results expresses their consistency. 

\begin{figure}[h!]
\begin{center}
\includegraphics[scale=.5]{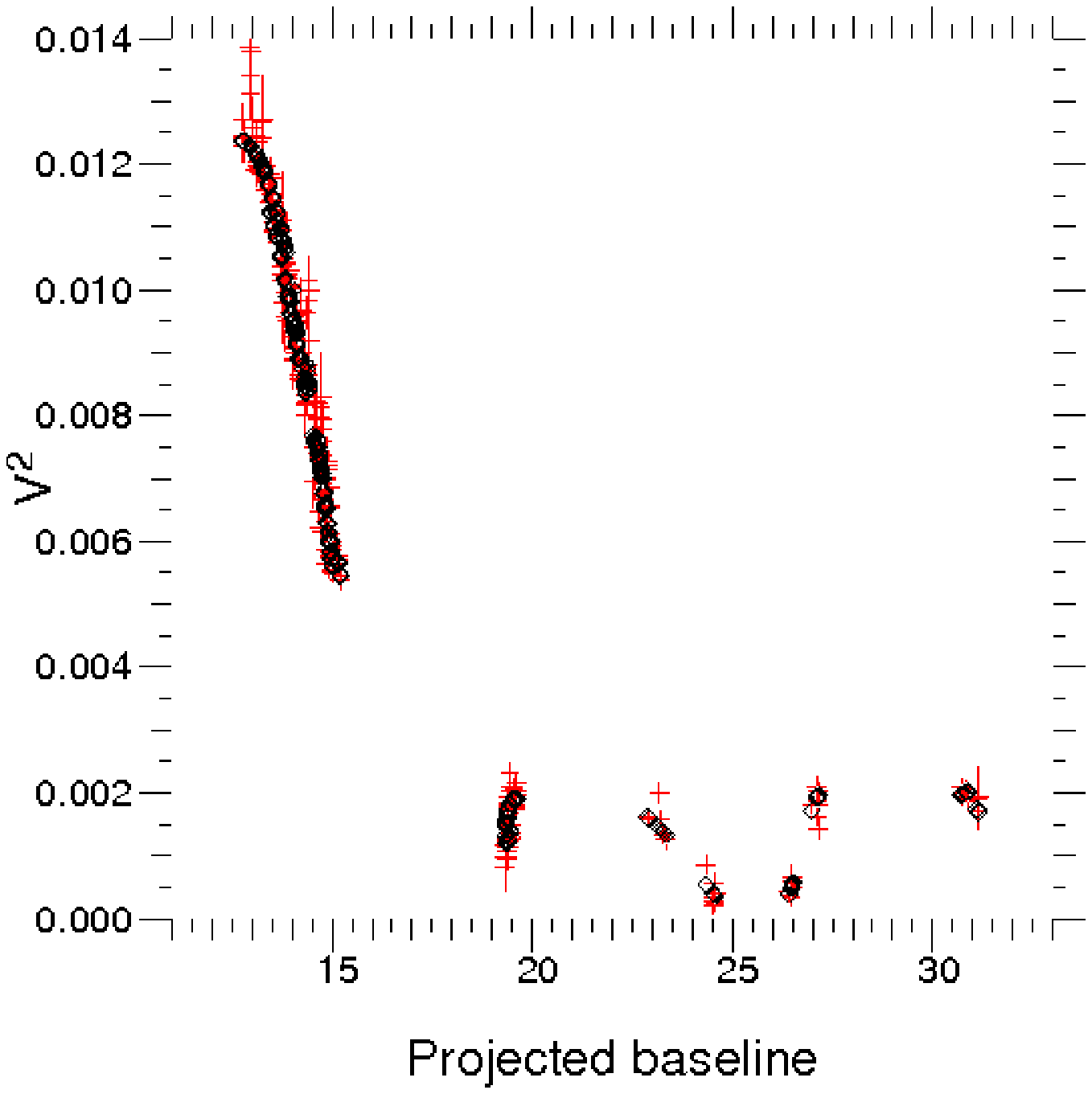}
\end{center}
\caption{\label{fig:2_spotsvis} Reconstructed squared visibilities from MIRA compared to the observed data. Only the second, the third and the fourth lobe are plotted here. We note that the third and fourth lobes are now well reproduced thanks to the contribution of two bright spots. The WISARD reconstructed visibilities are very similar to the MIRA ones and are not plotted here for conciseness.}
\end{figure}

We tried different regularization types and saw that the resulting images were not different from the quadratically regularized ones. We believe the reality and the location of asymmetries on the stellar disc of \object{Betelgeuse} can be trusted. The interferometer image of a point source was also reconstructed (Fig.~\ref{fig:recons}): the asymmetries imaged at the surface of \object{Betelgeuse} are not artefacts caused by the sparse uv plane coverage. A visibility plot comparing the data and the reconstructed visibilities of MIRA is presented in Fig.~\ref{fig:2_spotsvis}. The fit of the $V^2$ data is clearly very good and the best of those obtained in this paper.

\begin{table*}
\centering
\begin{tabular}{lcccccl}
\hline
Parameter & MIRA-T1 & WISARD-T1 & MIRA-T2 &  WISARD-T2  \\
\hline
RRA-cen &     2.3  $\pm$ 1.5 &  2.9   $\pm$ 1.5 & -1.3  $\pm$ 1.5  & -0.6  $\pm$ 1.5  \\
RDEC-cen & 7.5  $\pm$ 1.5 & 7.1  $\pm$ 1.5 &  - 4.5   $\pm$ 1.5 &  -4.7  $\pm$ 1.5  \\ 
FWHM-RA &   10.32   $\pm$ 2 & 11.08  $\pm$ 2 &  6.86     $\pm$ 2 &  8.89  $\pm$ 2   \\ 
FWHM-DEC &   5.80 $\pm$ 2 & 6.38  $\pm$ 2 &     4.32  $\pm$ 2 &   5.65 $\pm$ 2  \\ 
Amplitude   &         8     $\pm$  1 &   9    $\pm$ 1  &    4 $\pm$ 1 &         5   $\pm$ 1   \\
(\% the max) & & & & \\
     \hline

\end{tabular}
\caption{\label{tab:2_spots}Parameters of a Gaussian fit to the two bright spots revealed by MIRA and WISARD. T1 corresponds to the northern spot, T2 to the southern one.}
\end{table*}




In Table~\ref{tab:2_spots}, we see that  T1 is bigger than T2 and  brighter by a factor of 2. T1 seems more elongated in the northwestern direction than T2 though the inclination changes between the two images. The limiting resolution we achieved during our observations is about 11 mas. Separated by a distance of about 12.5 mas, the asymmetry created by an unequally bright  pair of spots such as T1 and T2 was resolved by the interferometer. 
With an angular size of about 11 mas, T1 is resolved and we can trust the presence of this spot at this location on the stellar surface. T1 might also be a combination of smaller scale structures. However it is not the case for T2 which is smaller than the maximum resolution we got. We therefore cannot conclude on the size that T2 may intrinsically have. \object{Betelgeuse} is certainly more complex than it appears on these images. Particularly the presence of structure whose size is smaller than 10 mas is very likely when looking at these images. However, the respective reduced $\chi^{2}$ values of MIRA and WISARD are low enough that it can be concluded that the reconstructed images (respectively 5.7 and 4.7) represent a fairly convincing modeling of the star surface given the accuracy, the complexity and the amount of data.


\subsection{Parametric fit of two spots}
\begin{figure}[h!]
\begin{center}
\includegraphics[scale=.5]{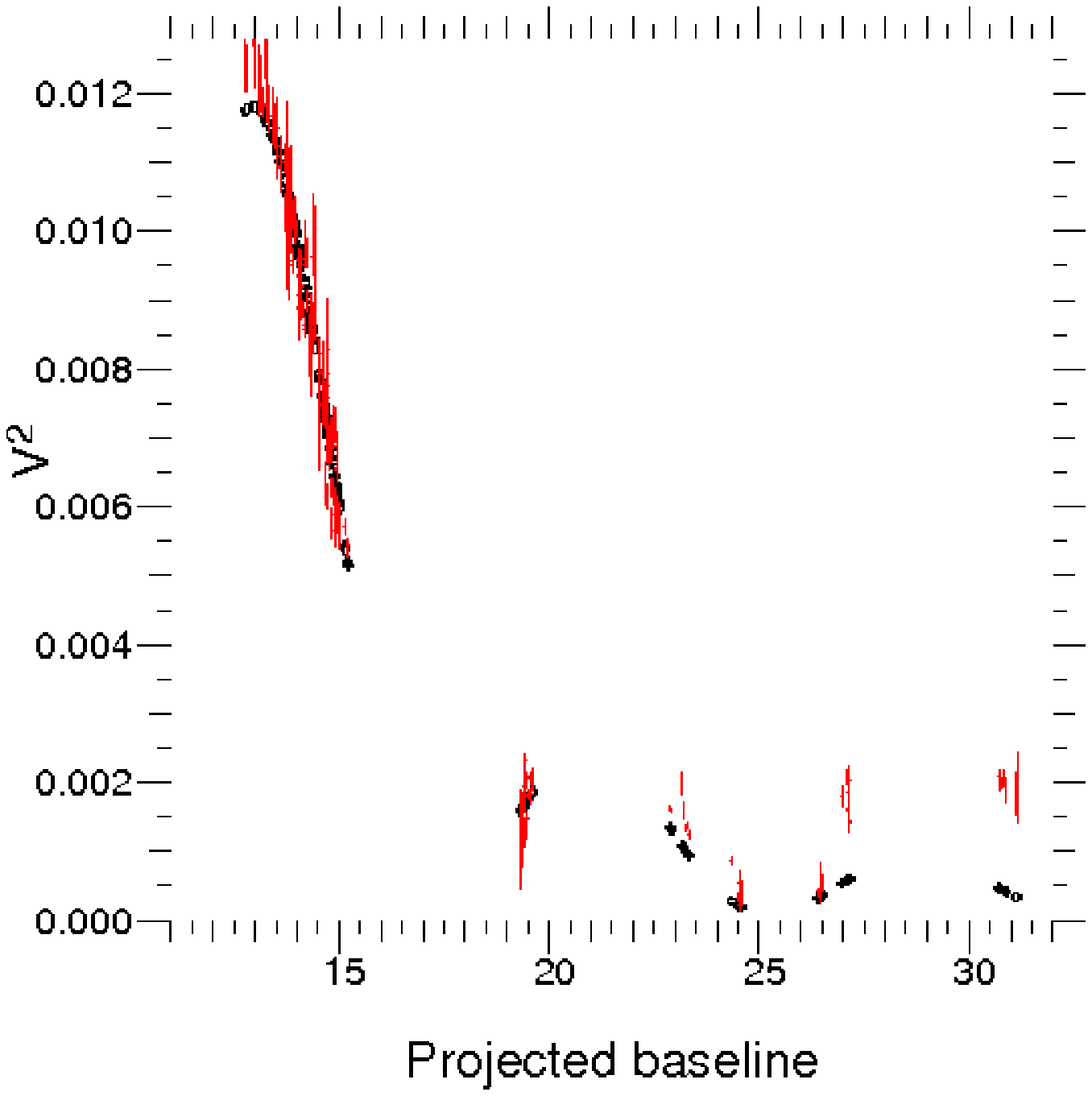}
\includegraphics[scale=.5]{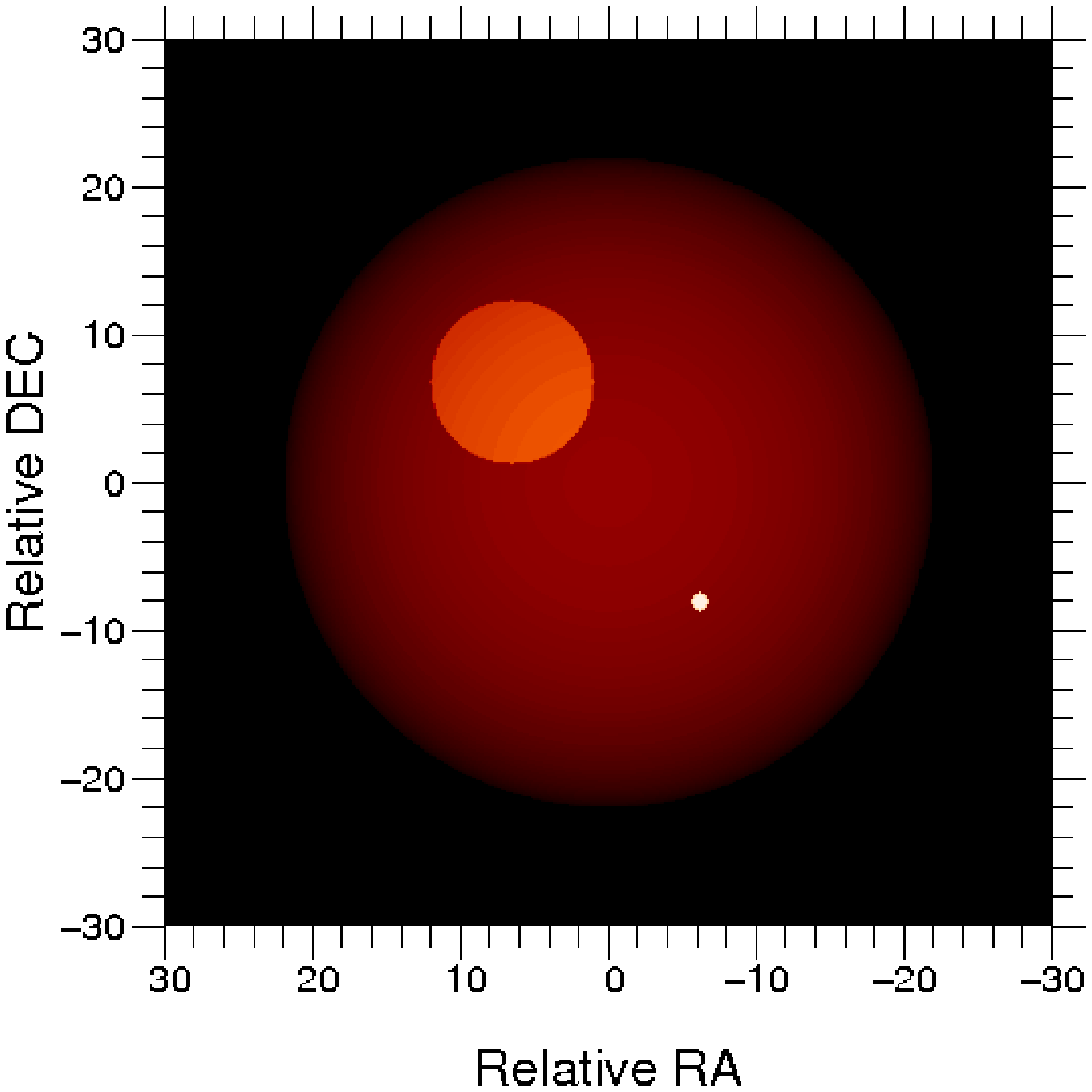}
\end{center}
\caption{\label{fig:2taches_spot}\textbf{Upper panel}: Fit of the squared visibilities by a limb-darkened disk plus 2 spots. Only the second, the third and the fourth lobe are plotted here. We note that the fourth lobe is poorly reproduced. \textbf{Lower panel}: Parametric image achieved by fitting visbilities and closure phases.}
\end{figure}

Image reconstruction has unveiled unforeseen structures with the use of an a priori image and some regularization conditions. Nevertheless, fitting a parametric model is an important task since it models the object with a limited number of parameters without the influence of the regularization. As the two different regularization methods point towards the existence of two bright structures near the center of the star, we used this information to simultaneously model visibilities and closure phases. We thus modeled two spots as uniform disks plus the limb-darkened disk and environment derived in section~\ref{LDsect}. The diameter of the northern spot was set to the value found in the previous section. The free parameters were the positions and relative fluxes of the two spots and the diameter of the southern spot. Results are shown in Fig.~\ref{fig:2taches_spot} and in Table~\ref{tab:2_spots}.
 
The best fit gives a reduced $\chi^{2}$ of 15 for the spot positions given in Table~\ref{tab:2_spots}.
The flux ratio and diameters corresponding to the best fit are listed in Table~\ref{tab:2_spots2}.

 \begin{table}[h!]
\centering
\begin{tabular}{lcccl}
\hline
Parameter & Param-T1 & Param-T2 \\
\hline
RRA-cen &     6.5 $\pm$ 1.5 &  -6.2 $\pm$ 1.5  \\
RDEC-cen & 6.8 $\pm$ 1.5 & -8.0 $\pm$ 1.5   \\ 
UD diameter in mas &   11  & 0.02 $\pm$ 1  \\ 
Amplitude   &         2  $\pm$ 1 &   2 $\pm$ 1   \\
(\% the max) & & & & \\
     \hline

\end{tabular}
\caption{\label{tab:2_spots2}Parameters of the 2 spots found by the parametric fit. Spots are modeled as uniform disks.}
\end{table}

The third visibility lobe fit is clearly improved with a two-spot model compared to circularly symmetric models. It is confirmed that the southern spot is not resolved although the flux ratios found by this method do not match those derived from the regularized images. Contrary to the case of the regularized images, the fourth lobe is not well reproduced by the parametric fit, suggesting that even smaller scale bright structures are needed to interpret these data. This may partially explain the discrepancies with the regularized images which better fit the data. More generally, all the structures disclosed by reconstruction algorithms are very difficult to model and fit in a parametric way because of the number of free parameters required. Although the 2-spot model catches most of \object{Betelgeuse} surface brightness complexity and can be considered a fair description of the object spatial brightness distribution, going further requires even smaller and fainter details (as images in Fig.~\ref{fig:recons} show) thus probably showing the complexity of the star surface.




    \section{Conclusion}
    \label{conclusect} 
    In this paper, we report on high accuracy interferometric measurements of \object{Betelgeuse} obtained with IOTA/IONIC up to a resolution of $\sim$11 mas. The precision on squared visibilities enabled by the use of guided optics was better than a percent. The complexity revealed by the data led us to use different methods and models which gave important results. We: 
    \begin{itemize}
    \item{determine the limb-darkened diameter of \object{Betelgeuse} using parametric and MARCS models as well as a new effective temperature,}

    \item{detect a fully resolved circumstellar environment as bright as 4\% of the total flux. It is very likely that the silicate dust shell is a significant contributor to this environment, and}
  
\item{image two spots at the surface of the star. One of the spots is unresolved and the other one is as large as one half of the stellar radius.}
    \end{itemize}

    
No MOLsphere has been detected in agreement with previous work at near-infrared wavelengths. We thus find a compatibility with the predictions published in \cite{1975ApJ...195..137S} about the existence of large structures on the surface of \object{Betelgeuse} due to convection. As \cite{2009arXiv0902.2602K} observed for other red supergiants, we note that these structures are responsible for up to $\sim$10 percent of the total flux. 

If one assumes that \object{Betelgeuse} and the spot T1 radiate like blackbodies,  an effective temperature of 3600 K for \object{Betelgeuse} leads to a spot T1 effective temperature of 4125 K (11 mas diameter and a flux ratio of 0.085). This is compatible with \cite{1975ApJ...195..137S} who predicted up to 1000 K temperature difference between the hottest and coolest parts of a convective surface. If these spotty structures are similar to those observed on the Sun, then they could be faculae which are bright zones surrounding dark convection cells. Their presence on \object{Betelgeuse} would show that a magnetic field plays an important role as it is the case for the Sun.



 To go further, it seems that the analysis of this dataset could be improved by injecting small scale ($<$ 10 mas) structures in the modeling. This can be done by comparing our data to simulated visibilities and closure phases generated from RSG hydro-radiative simulations \citep[such as presented in][]{2009arXiv0907.1860C}. This would undoubtedly bring new clues for the convective nature of the structures found in the present work. The convective motions could play an important role in the mass-loss process operating in RSGs, because of the strong decrease in effective gravity due to turbulent pressure \citep{2007A&A...469..671J}.
The bright plume found by \cite{2009arXiv0907.1843K} might be the first direct visible evidence of a mass loss mechanism in RSGs. Its formation could be explained by large upwards moving convective cell or by rotation, through the presence of a hot spot at the location of the polar cap of the star. The possibility that this plume is connected to a spot on \object{Betelgeuse} should be investigated with interferometric techniques in the near future. These perspectives hold the promise of determining a first estimation of the convection scale in RSGs which would represent a new step in the understanding of  the mass loss enigma in RSGs.




   \begin{acknowledgements}
   This work was supported by a grant from R{\'e}gion Ile-de-France. We also received the support of PHASE, the high angular resolution partnership between ONERA, Observatoire de Paris, CNRS and University Paris Diderot Paris.  X.H. thanks Denis Defr{\`e}re and Olivier Absil for fruitful discussions on the quality of IOTA measurements. STR acknowledges support by NASA grant NNH09AK731. TV acknowledges financial support from the FWO, Flanders. Part of this research was carried out at the Jet Propulsion Laboratory under a contract with the National Aeronautics and Space Administration. US Government sponsorship acknowledged.

   \end{acknowledgements}


\begin{thebibliography}{78}

\bibitem[Aitken(2005)]{2005ASPC..343..293A} Aitken, D.~K.\ 2005, Astronomical Polarimetry: Current Status and Future Directions, 343, 293 

\bibitem[Baldwin et al.(1996)]{1996A&A...306L..13B} Baldwin, J.~E., et al.\ 1996, \aap, 306, L13 

\bibitem[Bedding et al.(1997)]{1997MNRAS.286..957B} Bedding, T.~R., Zijlstra, A.~A., von der Luhe, O., Robertson, J.~G., Marson, R.~G., Barton, J.~R., \& Carter, B.~S.\ 1997, \mnras, 286, 957 

\bibitem[{{Bester} {et~al.}(1991){Bester}, {Danchi}, {Degiacomi}, {Townes}, \&  {Geballe}}]{Bester1991}{Bester}, M., {Danchi}, W.~C., {Degiacomi}, C.~G., {Townes}, C.~H., \&  {Geballe}, T.~R. 1991, \apjl, 367, L27

\bibitem[Berger et al.(2003)]{2003SPIE.4838.1099B} Berger, J.-P., et al.\ 2003, \procspie, 4838, 1099 

\bibitem[Bernat et al.(1979)]{1979ApJ...233L.135B} Bernat, A.~P., Hall, D.~N.~B., Hinkle, K.~H., \& Ridgway, S.~T.\ 1979, \apjl, 233, L135 

\bibitem[{{Bouwman}(2001)}]{Bouwman2001}{Bouwman}, J. 2001, PhD thesis, University of Amsterdam

\bibitem[{{Bouwman} {et~al.}(2000){Bouwman}, {de Koter}, {van den Ancker}, \&  {Waters}}]{Bouwman2000}{Bouwman}, J., {de Koter}, A., {van den Ancker}, M.~E., \& {Waters}, L.~B.~F.~M. 2000, \aap, 360, 213

\bibitem[Buscher et al.(1990)]{1990MNRAS.245P...7B} Buscher, D.~F., Baldwin, J.~E., Warner, P.~J., \& Haniff, C.~A.\ 1990, \mnras, 245, 7P 

\bibitem[Brunish \& Truran(1982)]{1982ApJ...256..247B} Brunish, W.~M., \& Truran, J.~W.\ 1982, \apj, 256, 247 

\bibitem[Castor(1993)]{1993ASPC...35..297C} Castor, J.~I.\ 1993, Massive Stars:  Their Lives in the Interstellar Medium, 35, 297 

\bibitem[Chiavassa et al.(2009a)]{2009arXiv0907.1860C} Chiavassa, A., Plez, B., Josselin, E., \& Freytag, B.\ 2009a, arXiv:0907.1860 

\bibitem[Chiavassa et al.(2009b)]{Chiavassa et al. 2010} Chiavassa, A., Haubois, X., Plez, B., Perrin, G., Josselin, E., \& Freytag, B., 2009b, in preparation

\bibitem[Claret(2004)]{2004A&A...428.1001C} Claret, A.\ 2004, \aap, 428, 1001 

\bibitem[Cornwell \& Wilkinson(1981)]{1981MNRAS.196.1067C} Cornwell, T.~J., \& Wilkinson, P.~N.\ 1981, \mnras, 196, 1067 

\bibitem[Coud{\'e} du Foresto et al.(1997)]{1997A&AS..121..379C} Coud{\'e} du Foresto, V., Ridgway, S., \& Mariotti, J.-M.\ 1997, \aaps, 121, 379 

\bibitem[Coud{\'e} du Foresto et al.(1998)]{1998SPIE.3350..856C} Coud{\'e} du Foresto, V., Perrin, G., Ruilier, C., Mennesson, B.~P., Traub, W.~A., \& Lacasse, M.~G.\ 1998, \procspie, 3350, 856 

\bibitem[{{Danchi} {et~al.}(1994){Danchi}, {Bester}, {Degiacomi}, {Greenhill},  \& {Townes}}]{Danchi1994}{Danchi}, W.~C., {Bester}, M., {Degiacomi}, C.~G., {Greenhill}, L.~J., \&  {Townes}, C.~H. 1994, \aj, 107, 1469

\bibitem[Danielson et al.(1965)]{1965ApJ...141..116D} Danielson, R.~E., Woolf, N.~J., \& Gaustad, J.~E.\ 1965, \apj, 141, 116 

\bibitem[Dravins(1982)]{1982ARA&A..20...61D} Dravins, D.\ 1982, \araa, 20, 61 

\bibitem[Dyck et al.(1992)]{1992AJ....104.1982D} Dyck, H.~M., Benson, J.~A., Ridgway, S.~T., \& Dixon, D.~J.\ 1992, \aj, 104, 1982 

\bibitem[{{Feautrier}(1964)}]{Feautrier1964}{Feautrier}, P. 1964, C.R.Acad.Sc.Paris, 258, 3189

\bibitem[Freytag \& H{\"o}fner(2008)]{2008A&A...483..571F} Freytag, B., H{\"o}fner, S.\ 2008, \aap, 483, 571 

\bibitem[Freytag et al.(2002)]{2002AN....323..213F} Freytag, B., Steffen, M., \& Dorch, B.\ 2002, Astronomische Nachrichten, 323, 213 

\bibitem[de Jager(1984)]{1984A&A...138..246D} de Jager, C.\ 1984, \aap, 138, 246 

\bibitem[de Jager et al.(1988)]{1988A&AS...72..259D} de Jager, C., Nieuwenhuijzen, H., \& van der Hucht, K.~A.\ 1988,\aaps, 72, 259 

\bibitem[Gilliland \& Dupree(1996)]{1996ApJ...463L..29G} Gilliland, R.~L., \& Dupree, A.~K.\ 1996, \apjl, 463, L29 

\bibitem[Guinan(1984)]{1984LNP...193..336G} Guinan, E.~F.\ 1984, Cool Stars, Stellar Systems, and the Sun, 193, 336 

\bibitem[Gustafsson et al.(1975)]{1975A&A....42..407G} Gustafsson, B., Bell, R.~A., Eriksson, K., \& Nordlund, A.\ 1975, \aap, 42,407 

\bibitem[van Hamme(1993)]{1993AJ....106.2096V} van Hamme, W.\ 1993, \aj, 106, 2096 

\bibitem[{{Harper} {et~al.}(2001){Harper}, {Brown}, \& {Lim}}]{Harper2001}{Harper}, G.~M., {Brown}, A., \& {Lim}, J. 2001, \apj, 551, 1073

\bibitem[Harper et al.(2008)]{2008AJ....135.1430H} Harper, G.~M., Brown, A., \& Guinan, E.~F.\ 2008, \aj, 135, 1430 

\bibitem[Harper et al.(2009)]{2009AIPC.1094..868H} Harper, G.~M., Carpenter, K.~G., Ryde, N., Smith, N., Brown, J., Brown, A., \& Hinkle, K.~H.\ 2009, American Institute of Physics Conference Series, 1094, 868 

\bibitem[Hestroffer(1997)]{1997A&A...327..199H} Hestroffer, D.\ 1997, \aap, 327, 199 

\bibitem[Hill \& Willson(1979)]{1979ApJ...229.1029H} Hill, S.~J., \& Willson, L.~A.\ 1979, \apj, 229, 1029 

\bibitem[Hinkle et al.(1982)]{1982ApJ...252..697H} Hinkle, K.~H., Hall, D.~N.~B., \& Ridgway, S.~T.\ 1982, \apj, 252, 697 

\bibitem[{{Howell} {et~al.}(1981){Howell}, {McCarthy}, \& {Low}}]{Howell1981}{Howell}, R.~R., {McCarthy}, D.~W., \& {Low}, F.~J. 1981, \apjl, 251, L21

\bibitem[Humphreys \& Davidson(1979)]{1979ApJ...232..409H} Humphreys, R.~M., \& Davidson, K.\ 1979, \apj, 232, 409 

\bibitem[Humphreys(1983)]{1983ApJ...269..335H} Humphreys, R.~M.\ 1983, \apj, 269, 335

\bibitem[Humphreys(1991)]{1991IAUS..143..485H} Humphreys, R.~M.\ 1991, Wolf-Rayet Stars and Interrelations with Other Massive Stars in Galaxies, 143, 485 

\bibitem[Josselin \& Plez(2007)]{2007A&A...469..671J} Josselin, E., \& Plez, B.\ 2007, \aap, 469, 671 

\bibitem[Kervella et al.(2009)]{2009arXiv0907.1843K} Kervella, P., Verhoelst, T., Ridgway, S.~T., Perrin, G., Lacour, S., Cami, J., \& Haubois, X.\ 2009, arXiv:0907.1843 

\bibitem[Kiss et al.(2009)]{2009arXiv0902.2602K} Kiss, L.~L., Monnier, J.~D., Bedding, T.~R., Tuthill, P., Zhao, M., Ireland, M.~J., \& ten Brummelaar, T.~A.\ 2009, arXiv:0902.2602 

\bibitem[Lacour et al.(2008)]{2008A&A...485..561L} Lacour, S., et al.\ 2008, \aap, 485, 561 

\bibitem[Lambert \& Snell(1975)]{1975MNRAS.172..277L} Lambert, D.~L., \& Snell, R.~L.\ 1975, \mnras, 172, 277 

\bibitem[Lambert et al.(1984)]{1984ApJ...284..223L} Lambert, D.~L., Brown, J.~A., Hinkle, K.~H., \& Johnson, H.~R.\ 1984, \apj, 284, 223 


\bibitem[Langer et al.(1989)]{1989A&A...224L..17L} Langer, N., El Eid, M.~F., \& Baraffe, I.\ 1989, \aap, 224, L17 

\bibitem[Le Besnerais(2008)]{article_Mugnier}G. Le Besnerais, S. Lacour, L. M. Mugnier, E. Thi\'{e}baut, G. Perrin, and S. Meimon. Advanced imaging methods for long-baseline optical interferometry. IEEE Journal of Selected Topics in Signal Processing, 2, October 2008,5,767

\bibitem[Levesque et al.(2005)]{2005ApJ...628..973L} Levesque, E.~M., Massey, P., Olsen, K.~A.~G., Plez, B., Josselin, E., Maeder, A., \& Meynet, G.\ 2005, \apj, 628, 973 

\bibitem[Lim et al.(1998)]{1998Natur.392..575L} Lim, J., Carilli, C.~L., White, S.~M., Beasley, A.~J., \& Marson, R.~G.\ 1998, \nat, 392, 575 

\bibitem[Lobel \& Dupree(2001)]{Lobel2001} Lobel, A., \& Dupree, A.~K.\ 2001, \apj, 558, 815 

\bibitem[Manduca(1979)]{1979A&AS...36..411M} Manduca, A.\ 1979, \aaps, 36, 411 

\bibitem[Maeder \& Conti(1994)]{1994ARA&A..32..227M} Maeder, A., \& Conti, P.~S.\ 1994, \araa, 32, 227 

\bibitem[Meimon (2005)]{SergephD}Meimon, S., PhD Thesis Universit\'{e} Paris Sud, 2005

\bibitem[Meimon et al.(2005)]{2005JOSAA..22.2348M} Meimon, S., Mugnier, L.~M., \& Le Besnerais, G.\ 2005, Journal of the Optical Society of America A, 22, 2348 

\bibitem[Meimon et al.(2009)]{2008JOSAA..26..108M} Meimon, S., Mugnier, L.~M., \& Le Besnerais, G.\ 2009, Journal of the Optical Society of America A, 26, 108 

\bibitem[{{Mihalas}(1978)}]{Mihalas1978}{Mihalas}, D. 1978, {Stellar atmospheres /2nd edition/} (San Francisco, W.~H.~Freeman and Co., 1978.~650 p.)

\bibitem[Monnier(2001)]{2001PASP..113..639M} Monnier, J.~D.\ 2001, \pasp, 113, 639 

\bibitem[Monnier et al.(2004)]{2004ApJ...602L..57M} Monnier, J.~D., et al.\ 2004, \apjl, 602, L57 

\bibitem[Ohnaka(2004)]{2004A&A...421.1149O} Ohnaka, K.\ 2004, \aap, 421, 1149

\bibitem[Ohnaka et al.(2009)]{2009arXiv0906.4792O} Ohnaka, K., et al.\ 2009, arXiv:0906.4792 

 \bibitem[Pedretti et al.(2004)]{2004PASP..116..377P} Pedretti, E., et al.\ 2004, \pasp, 116, 377

\bibitem[Pedretti et al.(2005)]{2005ApOpt..44.5173P} Pedretti, E., et al.\ 2005, \ao, 44, 5173 

\bibitem[Perrin et al.(2007)]{2007A&A...474..599P} Perrin, G., et al.\ 2007, \aap, 474, 599 

\bibitem[Perrin et al.(2004)]{2004A&A...418..675P} Perrin, G., Ridgway, S.~T., Coud{\'e} du Foresto, V., Mennesson, B., Traub, W.~A., \& Lacasse, M.~G.\ 2004, \aap, 418, 675 

\bibitem[Perrin(2003)]{2003A&A...398..385P} Perrin, G.\ 2003, \aap, 398, 385 

\bibitem[Perrin et al.(1998)]{1998IAUS..189P..18P} Perrin, G., Coud{\'e} du Foresto, V., Ridgway, S.~T., Mariotti, J.-M., Carleton, N.~P., \& Traub, W.~T.\ 1998, Fundamental Stellar Properties, 189, 18P 

\bibitem[Plez et al.(1992)]{1992A&A...256..551P} Plez, B., Brett, J.~M., \& Nordlund, A.\ 1992, \aap, 256, 551 

\bibitem[Plez et al.(1993)]{1993ApJ...418..812P} Plez, B., Smith, V.~V., \& Lambert, D.~L.\ 1993, \apj, 418, 812 

\bibitem[Ryde et al.(2006)]{2006ApJ...637.1040R} Ryde, N., Harper, G.~M., Richter, M.~J., Greathouse, T.~K., \& Lacy, J.~H.\ 2006, \apj, 637, 1040 

\bibitem[Schwarzschild(1975)]{1975ApJ...195..137S} Schwarzschild, M.\ 1975, \apj, 195, 137 

\bibitem[{{Skinner} {et~al.}(1997){Skinner}, {Dougherty}, {Meixner}, {Bode},  {Davis}, {Drake}, {Arens}, \& {Jernigan}}]{Skinner1997}{Skinner}, C.~J., {Dougherty}, S.~M., {Meixner}, M., {et~al.} 1997, \mnras,  288, 295

\bibitem[{{Sloan} {et~al.}(1993){Sloan}, {Grasdalen}, \& {Levan}}]{Sloan1993}{Sloan}, G.~C., {Grasdalen}, G.~L., \& {Levan}, P.~D. 1993, \apj, 404, 328

\bibitem[Soker \& Kastner(2003)]{2003ApJ...592..498S} Soker, N., \& Kastner, J.~H.\ 2003, \apj, 592, 498 

\bibitem[Sutton et al.(1977)]{Sutton} Sutton, E.~C., Storey, J.~W.~V., Betz, A.~L., Townes, C.~H., \& Spears, D.~L.\ 1977, \apjl, 217, L97 

\bibitem[Tatebe et al.(2007)]{2007ApJ...670L..21T} Tatebe, K., Chandler, A.~A., Wishnow, E.~H., Hale, D.~D.~S., \& Townes, C.~H.\ 2007, \apjl, 670, L21 

\bibitem[Thi{\'e}baut(2002)]{2002SPIE.4847..174T} Thi{\'e}baut, E.\ 2002, \procspie, 4847, 174 

\bibitem[Thi{\'e}baut(2008)]{2008SPIE.7013E..43T} Thi{\'e}baut, E.\ 2008, \procspie, 7013,  

\bibitem[Townes et al.(2009)]{2009ApJ...697L.127T} Townes, C.~H., Wishnow, E.~H., Hale, D.~D.~S., \& Walp, B.\ 2009, \apjl, 697, L127 

\bibitem[Traub et al.(2003)]{2003SPIE.4838...45T} Traub, W.~A., et al.\ 2003, \procspie, 4838, 45 

\bibitem[Tsuji(1988)]{988A&A...197..185T} Tsuji, T.\ 1988, \aap, 197, 185 

\bibitem[Tsuji(2000a)]{2000ApJ...538..801T} Tsuji, T.\ 2000a, \apj, 538, 801 

\bibitem[Tsuji(2000b)]{2000ApJ...540L..99T} Tsuji, T.\ 2000b, \apjl, 540, L99

\bibitem[Uitenbroek et al.(1998)]{1998AJ....116.2501U} Uitenbroek, H., Dupree, A.~K., \& Gilliland, R.~L.\ 1998, \aj, 116, 2501 

\bibitem[{{Verhoelst} {et~al.}(2006){Verhoelst}, {Decin}, {van Malderen},  {Hony}, {Cami}, {Eriksson}, {Perrin}, {Deroo}, {Vandenbussche}, \&  {Waters}}]{Verhoelst2006}{Verhoelst}, T., {Decin}, L., {van Malderen}, R., {et~al.} 2006, \aap, 447, 311

\bibitem[Weiner(2004)]{2004ApJ...611L..37W} Weiner, J.\ 2004, \apjl, 611, L37 

\bibitem[Willson(2000)]{2000ARA&A..38..573W} Willson, L.~A.\ 2000, \araa, 38, 573 

\bibitem[Wilson et al.(1997)]{1997MNRAS.291..819W} Wilson, R.~W., Dhillon, V.~S., \& Haniff, C.~A.\ 1997, \mnras, 291, 819 

\bibitem[Wilson et al.(1992)]{1992MNRAS.257..369W} Wilson, R.~W., Baldwin, J.~E., Buscher, D.~F., \& Warner, P.~J.\ 1992, \mnras, 257, 369 

\bibitem[Wing \& Spinrad(1970)]{1970ApJ...159..973W} Wing, R.~F., \& Spinrad, H.\ 1970, \apj, 159, 973 

\bibitem[Wood(1979)]{1979ApJ...227..220W} Wood, P.~R.\ 1979, \apj, 227, 220 

\bibitem[Woosley \& Weaver(1986)]{1986ARA&A..24..205W} Woosley, S.~E., \& Weaver, T.~A.\ 1986, \araa, 24, 205 

\bibitem[Young et al.(2000)]{2000MNRAS.315..635Y} Young, J.~S., et al.\ 2000, \mnras, 315, 635 

\bibitem[Zhao et al.(2007)]{2007ApJ...659..626Z} Zhao, M., et al.\ 2007, \apj, 659, 626 

\end{thebibliography}
\end{document}